\RequirePackage{desc-tex/styles/docswitch}

\documentclass[]{emulateapj}

\usepackage{desc-tex/styles/lsstdesc_macros}
\usepackage{graphicx}
\usepackage{hyperref}
\usepackage{ae, aecompl}
\usepackage[english]{babel}
\usepackage{amsmath,amstext}
\usepackage{fontawesome}
\usepackage{color, colortbl}
\usepackage{booktabs, tabularx, multirow, bbm}
\usepackage{xcolor}
\usepackage{fancyvrb}

\graphicspath{{./}{./figures/}}

\begin{document}

\title{Large-Scale Gravitational Lens Modeling with Bayesian Neural Networks for Accurate and Precise Inference of the Hubble Constant}

\author{{Ji~Won~Park} \altaffilmark{1, 2}, {Sebastian Wagner-Carena}\altaffilmark{1}, {Simon~Birrer}\altaffilmark{1}, {Philip~J.~Marshall}\altaffilmark{1, 2},
{Joshua Yao-Yu Lin}\altaffilmark{3},
{Aaron~Roodman}\altaffilmark{1, 2} (The LSST Dark Energy Science Collaboration)}
\altaffiltext{1}{Kavli Institute for Particle Astrophysics and Cosmology, Department of Physics, Stanford University, Stanford, CA, 94305}
\altaffiltext{2}{SLAC National Accelerator Laboratory, Menlo Park, CA, 94025}
\altaffiltext{3}{University of Illinois at Urbana-Champaign, Champaign, IL, 61820}


\begin{abstract}
  We investigate the use of approximate Bayesian neural networks (BNNs) in modeling hundreds of time-delay gravitational lenses for Hubble constant ($H_0$) determination. Our BNN was trained on synthetic HST-quality images of strongly lensed active galactic nuclei (AGN) with lens galaxy light included. The BNN can accurately characterize the posterior PDFs of model parameters governing the elliptical power-law mass profile in an external shear field. We then propagate the BNN-inferred posterior PDFs into ensemble $H_0$ inference, using simulated time delay measurements from a plausible dedicated monitoring campaign. Assuming well-measured time delays and a reasonable set of priors on the environment of the lens, we achieve a median precision of $9.3$\% per lens in the inferred $H_0$. A simple combination of 200 test-set lenses results in a precision of 0.5 $\textrm{km s}^{-1} \textrm{ Mpc}^{-1}$ ($0.7\%$), with no detectable bias in this $H_0$ recovery test. The computation time for the entire pipeline---including the training set generation, BNN training, and $H_0$ inference---translates to 9 minutes per lens on average for 200 lenses and converges to 6 minutes per lens as the sample size is increased. Being fully automated and efficient, our pipeline is a promising tool for exploring ensemble-level systematics in lens modeling for $H_0$ inference.
\end{abstract}



\section{Introduction}
\label{sec:intro}
The recent widening of the ``Hubble tension'' signals the need for rigorous tests of systematics in all cosmographic probes. The discrepancy in Hubble constant ($H_0$) measurements between early- and late-universe probes now lies at a $4 - 6\sigma$ level \citep{verde2019tensions}. Particularly valuable in this context are strong gravitational time delays -- observed when light from a variable source is lensed by a massive foreground object, creating multiple images with relative delays in photon arrival times \citep{refsdal1964possibility}. As time delay cosmography is fully independent of $H_0$ determination methods using the local distance ladder and the cosmic microwave background (CMB), it can serve as a check against sources of bias that may be affecting either method.

The {H0LiCOW} Collaboration inferred $H_0$ to 2.4\% precision using six lenses in the flat $\Lambda$CDM cosmology \citep{wong2019h0licow}. The uncertainty increases to 7\%, however, when the assumptions on the radial mass density profile of the lenses are relaxed and one additional lens is included \citep{birrer2020tdcosmohierarchical}. Further folding in the external information from 33 Sloan Lens ACS (SLACS) galaxy-galaxy lenses without time delays \citep{bolton2008sloan, auger2009sloan, shajib2020massive}, the precision improves to 5\% assuming that the deflector galaxies follow the same population statistics. According to \cite{birrer2020tdcosmostrategies}, a sample size of 40 time delay lenses and 200 galaxy-galaxy lenses can enable 1.2\% - 1.5\% precision necessary to resolve the $H_0$ tension.

The current modeling cycle in time delay cosmography does not scale well to the prospects of upcoming large-scale surveys. The Legacy Survey of Space and Time (LSST) at the Vera Rubin Observatory is expected to discover tens of thousands of lens systems, among them hundreds of lensed quasars \citep{collett2015population, oguri2010gravitationally}. To date, time delay cosmography has relied on a time-consuming and manual forward-modeling of observations. This approach takes several months under expert monitoring. With automation efforts, which are underway, the time may be reduced to several weeks \citep{shajib2019every}. 

The efficiency issues aside, the current method of fine-tuning each lens model on a case-by-case basis makes it difficult to conduct global sensitivity tests on the model assumptions. See \cite{shajib2019every} for a uniform forward-modeling of 13 quadruply lensed quasars (quads), among the first efforts to capitalize on the self-similarity of quads for automated (and thus consistent) lens modeling. A joint inference over hundreds of lenses requires a computationally efficient method with a uniform approach to modeling, so that systematics can be probed in an ensemble of lenses within reasonable time. 

Bayesian neural networks (BNNs) offer an efficient alternative to forward-modeling \citep{denker1991transforming}. They are a probabilistic variant of deep neural networks, which have demonstrated state-of-the-art performance in extracting highly abstract information from complex image data. \citet{hezaveh2017fast} and \citet{levasseur2017uncertainties} demonstrated the efficacy of BNNs in accurately and precisely characterizing the lens model parameter posterior probability density functions (PDFs) for individual lenses, assuming a singular isothermal ellipsoid (SIE) lens model. 
Not only do BNN-based methods preclude the need for human supervision, once trained, a BNN model can be applied to thousands of lens systems within seconds on a single GPU.

This paper connects the progress in BNN-based lens modeling to the $H_0$ inference stage, by extending BNN lens modeling to use all the features in a time delay lensed AGN system and combining the posterior PDFs from that modeling in an industry-standard joint inference of $H_0$ from a plausible near-future ensemble. 

We are guided by the following questions:
\begin{itemize}
    \item Can the BNN accurately characterize the individual lens model posterior PDFs, given our model assumptions? 
    \item If so, do the BNN-inferred lens model posterior PDFs enable unbiased $H_0$ recovery when propagated into a joint $H_0$ inference over 200 lenses?
    \item How sensitive are the $H_0$ predictions to the factors that are often considered when selecting lenses for follow-up -- namely, the exposure time, the lensed image configuration, and the Einstein ring brightness?
    \item Is our method efficient enough to handle large-scale tests of systematics? What is the net speed increase over traditional methods? 
\end{itemize}   

The goal of accurate and precise $H_0$ recovery places extra demands on lens modeling. This drives us to relax some of the assumptions made in the previous literature on lens models that are input to neural networks. We adopt the power-law elliptical lens mass distribution (PEMD) \citep{barkana1998fast}, a more complex form of the mass density profile than the fixed-slope SIE model used by \cite{hezaveh2017fast}, \cite{pearson2019use}, and \cite{schuldt2020holismokes} for their neural networks. PEMD is the model family currently used by the {H0LiCOW} and TDCOSMO collaborations in their time delay cosmography analyses \citep{wong2019h0licow, birrer2020tdcosmohierarchical}.

$H_0$ inference also requires precise source position recovery. As discussed in \cite{birrer2019astrometric}, the precision required on the source position is on the order of milliarcseconds. The ability of BNNs to constrain source positions to this level of precision has not yet been tested, but the accuracy of the predicted time delays (and hence the inferred $H_0$ value) will depend critically on this.

Lastly, we include the lens light in the images. In \cite{hezaveh2017fast}, the lens light was removed from the images via independent component analysis (ICA) before the images were passed into the neural network for training. \cite{pearson2019use} saw a 34\% reduction in accuracy of lens model recovery for images with lens light included, but report that multi-band imaging could alleviate the performance drop. In the present study we restrict ourselves to a single HST IR band and postpone investigation of multiple bands to further work.

In this paper, we demonstrate that BNN lens modeling successfully meets the above performance requirements defined by time delay cosmography. Given our model assumptions, BNNs can, in fact, characterize the posterior PDF with sufficient accuracy, so as to recover $H_0$ without bias from a joint 200-lens inference. The source code we developed for our work is also released for public use. The methodology and software presented in this paper are inherently versatile and allow extensions in many directions, including the hierarchical inference setup we developed in \cite{wagner2020hierarchical}. They promise to become core infrastructure in time delay cosmography, as the cosmology community prepares to beat down systematics for a large sample of lenses due to be available in a few years' time. 

This paper is organized as follows. 
Section \ref{sec:methods} details each step of our automated $H_0$ inference pipeline. In Section \ref{sec:results}, we report $H_0$ recovery results on 200 test-set lenses under varying noise levels, image configurations (double vs. quad), and Einstein ring brightness. Section \ref{sec:conclusions} situates our findings within a larger context of BNN-aided time delay cosmography and outlines next steps.

\section{Methods}
\label{sec:methods}
\begin{figure*}
\centering
\centering
\includegraphics[width=0.8\paperwidth]{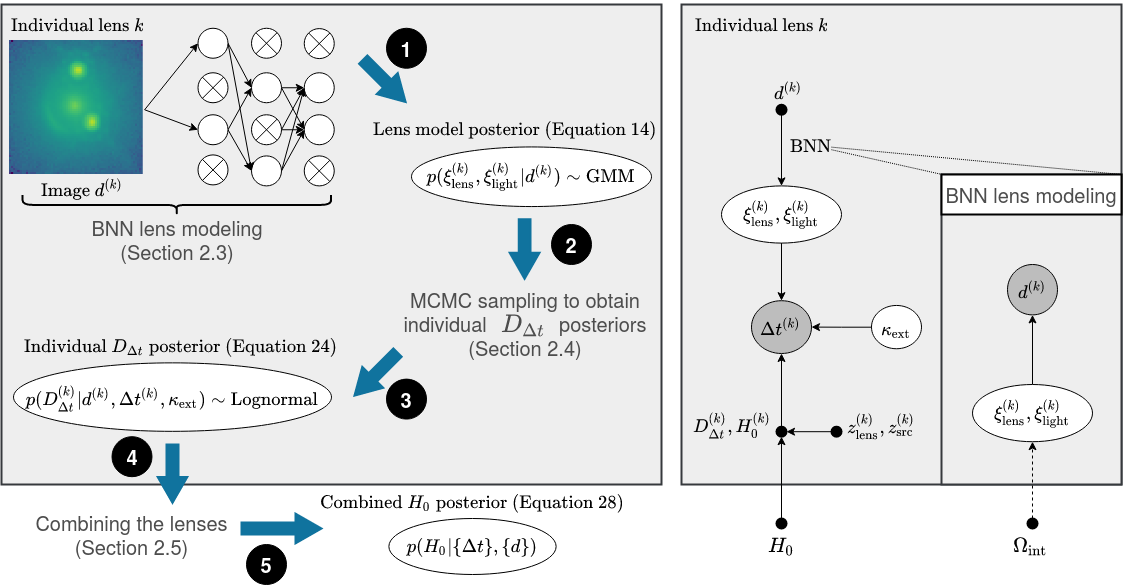}
\caption{Left: illustration of the $H_0$ inference pipeline in the form of a flowchart. Right: the dependence relation shown as a probabilistic graphical model (PGM). Dots refer to delta functions, or fixed values; shaded ovals refer to observed values, or data; and unshaded ovals refer to random variables.}
\label{fig:pipeline_diagram}
\end{figure*}

This section details the steps for constructing an $H_0$ inference pipeline with a BNN as the lens modeling engine, beginning with a brief theoretical background of time delay cosmography in Section \ref{sec:time_delay_cosmography}. In Section \ref{sec:data_and_models}, we state our assumptions about the lens population, instrument optics, observation conditions, and cosmology -- all of which we used to simulate the lensed AGN images and time delays.  Then, in Section \ref{sec:automated_lens_modeling}, we explain how the BNN models the individual lens model posteriors. The BNN-inferred lens model posterior becomes propagated into $H_0$ inference; Section \ref{sec:individual_h0_inference} describes this process on an individual lens level and Section \ref{sec:joint_h0_inference} on the joint sample level. The entire pipeline is illustrated in Figure \ref{fig:pipeline_diagram} as a flowchart and a probabilistic graphical model (PGM).

The implementation of the whole pipeline, including the BNN lens modeling and $H0$ inference, is available in the open-source Dark Energy Science Collaboration (DESC) Python package \textsc{H0rton}\footnote{\faicon{github} \url{http://github.com/jiwoncpark/h0rton}}. To generate the training set, we developed another DESC Python package \textsc{Baobab}\footnote{\faicon{github} \url{http://github.com/jiwoncpark/baobab}}, which wraps around the multi-purpose lens modeling package \textsc{Lenstronomy}\footnote{\faicon{github} \url{https://github.com/sibirrer/lenstronomy}} \citep{birrer2018lenstronomy} to render the images and compute the time delays. 

\subsection{Time delay cosmography} \label{sec:time_delay_cosmography}
Let us begin by reviewing the basic principles of time delay cosmography \citep{refsdal1964possibility}. Readers are referred to recent reviews, e.g. \cite{treu2016time}, for more details. When light rays from a background source are deflected by some foreground lens, the light travel time from the source to the observer depends on both their path length and the gravitational potential they must traverse. Assuming a single thin, isolated lens, the excess time delay of an image at position $\boldsymbol{\theta}$ originating from a source at position $\boldsymbol{\beta}$ relative to an unperturbed path is
\begin{align} \label{eq:excess_time_delays}
    t(\boldsymbol{\theta}, \boldsymbol{\beta}) = \frac{D_{\Delta t}}{c} \phi(\boldsymbol{\theta}, \boldsymbol{\beta})
\end{align}
where 
\begin{align} \label{eq:fermat_potential}
\phi(\boldsymbol{\theta}, \boldsymbol{\beta}) = \left[\frac{ (\boldsymbol{\theta} -\boldsymbol{\beta})^2}{2} - \psi(\boldsymbol{\theta})\right]
\end{align}
is the Fermat potential \citep{schneider1985new, blandford1986fermat} defined for the lensing potential $\psi(\boldsymbol{\theta})$ and $D_{\Delta t}$ is the time delay distance \citep{refsdal1964possibility, schneider1992gravitational, suyu2010dissecting}. The time delay distance is defined as
\begin{align}
    D_{\Delta t} \equiv (1 + z_{\rm lens})\frac{D_{\rm d} D_{\rm s}}{D_{\rm ds}}
\end{align}
where $z_{\rm lens}$ is the lens redshift and $D_{\rm d}, D_{\rm s}, D_{\rm ds}$ are the angular diameter distances from the observer to the lens, the observer to the source, and from the lens to the source, respectively. 

If the background source and the foreground lens are well aligned, we observe multiple images of the same background source. The position of the source with respect to the inner caustic determines the whether there are two images, making the lensing system a ``double'', or four images, making it a ``quad.'' The time delay between any pair of such lensed images is the difference of their excess time delays in Equation \ref{eq:excess_time_delays}:
\begin{align} \label{eq:relative_time_delays}
    \Delta t_{i, j} = \frac{D_{\Delta t}}{c} \left[ \phi(\boldsymbol{\theta}_i, \boldsymbol{\beta}) - \phi(\boldsymbol{\theta}_j, \boldsymbol{\beta})\right]
\end{align}
where $\boldsymbol{\theta}_i, \boldsymbol{\theta}_j$ are the positions of images $i, j$ in the image plane. 

If the source is variable, like an AGN, it is possible to measure the relative time delay $\Delta t_{i, j}$ by monitoring the fluxes of the images \citep{vanderriest1989value, schechter1997quadruple, fassnacht1999determination, kochanek2006time, courbin2011cosmograil}. The lensing potentials at the two image positions $\psi(\boldsymbol{\theta}_i), \psi(\boldsymbol{\theta}_j)$ and the source position $\boldsymbol{\beta}$ can be determined by lens modeling, yielding a model of the relative Fermat potential $\Delta \phi_{i, j}$. Given the measured relative time delay and the constrained relative Fermat potential, we can constrain the time-delay distance via
\begin{align} \label{eq:constraint_D_dt}
    D_{\Delta t} = \frac{c \Delta t_{i, j}}{\Delta \phi_{i, j}}.
\end{align}
Being inversely proportional to the absolute distance scale, $H_0$ scales with $D_{\Delta t}$ as
\begin{align}
    H_0 \propto D_{\Delta t}^{-1}.
\end{align}

\VerbatimFootnotes
\subsection{Simulated dataset and model assumptions}
\label{sec:data_and_models}
The BNN, our lens modeling tool, requires a large training set that spans the target parameter space with sufficient density. The training set is necessarily synthetic because (1) fewer than 100 lensed AGN have been discovered to date and (2) it defines the models we assume for the lens mass, lens light, and source profiles during the inference stage. Our training set consists of 512,000 images, and we validate and test on independent and identically distributed sets of 512 and 200 lenses, respectively. 

\begin{figure*}[!htb]
\centering
\includegraphics[width=1\textwidth]{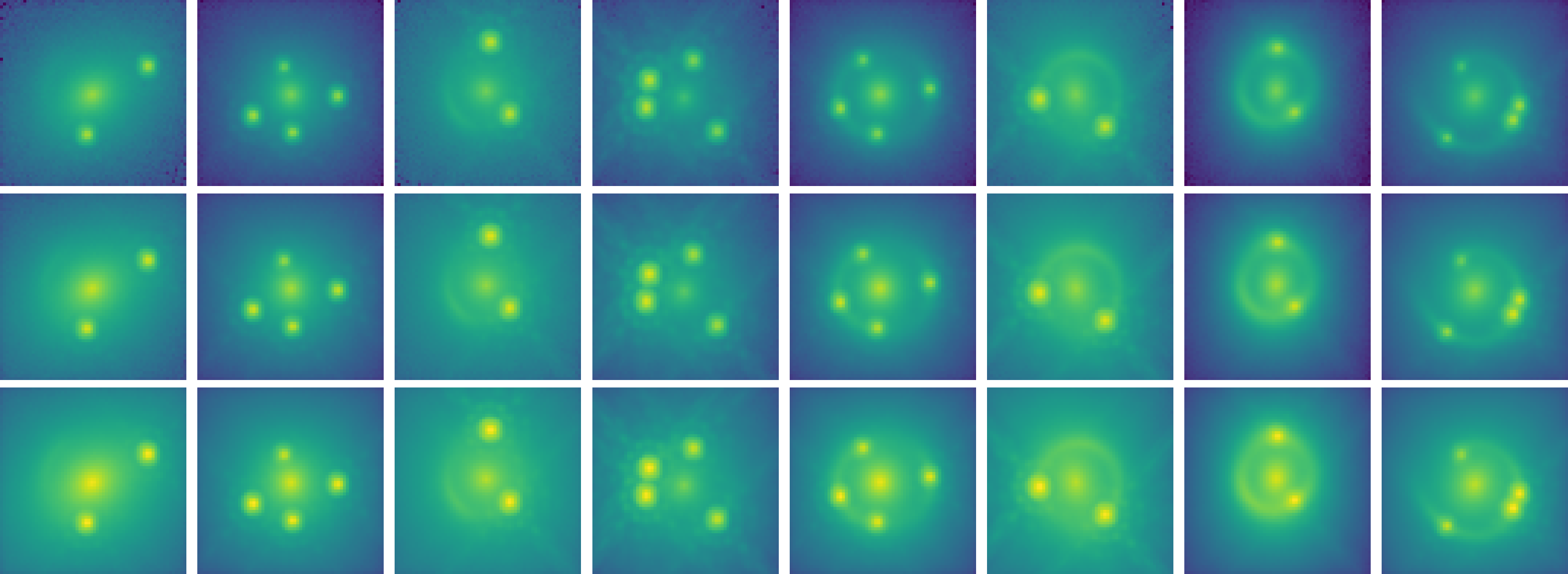}
\caption{Gallery of 8 images from our simulation, in log intensity scale. From the top row to the bottom, the exposure time varies as 0.5, 1, and 2 HST orbit(s). The columns are ordered such that, from left to right, the magnitude of the Einstein ring decreases. The 8 images have been sampled from the test set, but this also serves as a visualization of the training set, as the test-set and training-set images are drawn from the same distribution in our study. The same color scale is used across the exposure times, for each lens. }
\label{fig:training_set_images_full_transformed}
\end{figure*}

\subsubsection{Profile assumptions}
\label{sec:profile_assumptions}

This study requires model profiles that are flexible enough to describe plausible lensing systems well but not too complex, so as to allow for simple interpretations in the basic parameter recovery tests. Earlier work on neural network-based lens modeling had focused on the SIE lens mass profile \citep{hezaveh2017fast, levasseur2017uncertainties, schuldt2020holismokes, pearson2019use}. As an update to this model, we allow the 3D power-law mass slope $\gamma_{\rm lens}$ to vary by adopting the power-law elliptical lens mass distribution (PEMD) \citep{barkana1998fast}. Note that the precision in $\gamma_{\rm lens}$ roughly translates to the precision in $H_0$ so demonstrating that we can recover $\gamma_{\rm lens}$ is crucial. The PEMD profile can be written in terms of six parameters as
\begin{align}\label{eq:pemd}
    &\kappa(x, y) = \frac{3 - \gamma_{\rm lens}}{2} \left(\frac{\theta_E}{\sqrt{q_\textrm{lens} x^2 + y^2/q_\textrm{lens}}} \right)^{1 - \gamma_{\rm lens}} 
\end{align}
where $q_\textrm{lens}$ is the projected axis ratio and $\theta_E$ is the Einstein radius chosen such that it encloses the mean surface density in the spherical limit of $q_\textrm{lens} = 1$. The coordinates $(x, y)$ are the result of rotating the sky coordinates by the lens orientation angle $\phi_\textrm{lens}$, so that the $x$-axis and the major axis of the lens align, and then centering them at the lens position $(x_\textrm{lens}, y_\textrm{lens})$. We also include the external shear component, parameterized by the shear modulus $\gamma_{\rm ext}$ and the shear angle $\phi_{\rm ext}$. 

The lens galaxy light and the host galaxy light in our simulations follow the elliptical Sersic distribution, which can be expressed in terms of seven parameters as
\begin{align} \label{eq:sersic}
    &I(x, y) = I_* \exp \left[-k \bigg \{ \left(\frac{\sqrt{x^2 + y^2/{q_*}^2}}{R_*} \right)^{1/n} - 1 \bigg \} \right] 
\end{align}
where $I_e$ is the surface brightness amplitude at the half-light radius $R$, $k$ is a constant depending on the Sersic index $n$ such that $R$ encloses half of the light, and $q_*$ is the axis ratio. The coordinates $(x, y)$ are as defined for Equation \ref{eq:pemd}. As a simple approximation, we assume the lens light to share the centroid with the lens center. We parameterize the surface brightness amplitude $I_*$ in terms of the magnitude $m_*$ and convert into amplitude units using the instrument zeropoint in order to render the image.  

The AGN was modeled as an unresolved point source. To simulate microlensing, we added 10\% Gaussian errors to the magnifications of the lensed AGN images. 

The distribution of the model parameters in our training set serves as the implicit prior for our BNN. For the PEMD lens mass, external shear, and Sersic lens light, we chose parameter distributions slightly broader than those in the Time Delay Lens Modeling Challenge (TDLMC) \citep{ding2018time}. The distribution of the AGN host galaxy parameters was based on the estimates of source galaxy populations in galaxy-galaxy lenses presented in \cite{collett2015population}. See Table \ref{tab:parameter_dist} for the specific choice of hyperparameters defining the implicit prior.

When the true input model parameters were drawn from the implicit prior, the ellipticities were parameterized in terms of the axis ratio $q_{\rm lens}$ and complex orientation angle $\phi_{\rm lens}$ as defined above. Similarly, the external shear was parameterized in terms of the shear modulus $\gamma_{\rm ext}$ and complex shear angle $\phi_{\rm ext}$. But the 2$\pi$-periodic property of the angles introduces degeneracies in target space that makes the BNN prediction task ill-defined. For training the BNN, we thus parameterize the target lens mass ellipticity and external shear in terms of the coordinate values in their respective spaces:
\begin{align} \label{eq:ellip_shear_conversion}
& e_1 = \frac{1-q_{\text{lens}}}{1+q_{\text{lens}}} \cos (2 \phi_\text{lens}) \nonumber \\
& e_2 = \frac{1-q_{\text{lens}}}{1+q_{\text{lens}}} \sin (2 \phi_\text{lens}) \nonumber \\
& \gamma_1 = \gamma_\text{ext} \cos (2 \phi_\text{ext}) \nonumber \\
& \gamma_2 = \gamma_\text{ext} \sin (2 \phi_\text{ext})
\end{align}

Whereas there are known empirical covariances between subsets of our model parameters, such as a positive correlation between the ellipticities of lens mass and lens light, we assume the parameters to be \textit{a priori} independent. This has the effect of reducing the efficiency of our training set by including some less plausible lenses in the BNN training. If the trained BNN were to be tested on real data, there would indeed be greater motivation to encode some covariance in the training set. The independence assumption is a safe choice for the purposes of our study, however, as it prevents the BNN from relying on the prescribed covariances when it generates its predictions. Even in applications when a more realistic training set is necessary, one should exercise caution when encoding the covariances; implicit priors that are too tight can introduce bias in the BNN parameter inference as well as hierarchical inference. In fact, as we have demonstrated in \cite{wagner2020hierarchical}, broad implicit priors are generally advised because the wide support aids in numerical stability when performing importance sampling for hierarchical inference. 

 Throughout this paper, all magnitudes are given in the AB system with the WFC3/F160W filter zeropoint of 25.9463.

\subsubsection{Instrument and observation conditions}
\label{sec:instrument_observation}
We simulated images obtained with the Hubble Space Telescope (HST) using the Wide Field Camera 3 (WFC3) IR channel in the F160W band, following the design of TDLMC.  Dust extinction was not included, as it only has a weak effect in this filter. For simplicity, we approximated the PSF drizzling process by convolving the unconvolved image with the drizzled HST PSF template provided as part of the TDLMC Rung 1. The effective pixel size of this drizzled PSF was $0\farcs08$ and we fixed the image size to $64 \times 64$ pixels. The PSF FWHM was in the range $0\farcs14 \sim 0\farcs16$. 

The PSF-convolved, noiseless images were stored, so that noise could be added on the fly during training and testing. This setup exposes the network to different noise realizations of the same underlying system, which is known to help with generalization. It also precludes the need to generate new images for different noise levels. During training and testing, the stored noiseless images were scaled appropriately to simulate a new exposure time and \textsc{Baobab} efficiently computed the noise map on the GPU. The noise model included the background, readout, and Poisson CCD noise. We used the read noise of 4 $e^-$ and CCD gain of 2.5 $e^-$/ADU, following the mean instrument statistics reported for WFC3/IR F160W \citep{wfc3}. The sky brightness was calculated to be 22 mag/arcsec$^2$ based on the zodiacal light estimation in \cite{giavalisco2002new}, for the effective F160W filter wavelength of 1526.91nm. The median signal-to-noise ratios (SNR) for the 0.5, 1, and 2 HST orbits were 4, 9, and 20, respectively, where signal was taken to be the sum of the pixels of the image with the lens light subtracted. Figure \ref{fig:training_set_images_full_transformed} displays images in the training set, with a range of exposure times and Einstein ring brightness.

\subsubsection{Assumptions beyond the images}
\label{sec:additional_assumptions}
The lens model parameters can be constrained from the simulated imaging observables alone, but in order to perform cosmological inference, we need to assign additional information to each lensing system: the redshifts, density of matter in the environment, and the time delay measurements. 

The lens and source redshifts were drawn independently from Gaussian distributions centered at 0.5 and 2, respectively, as presented in Table \ref{tab:parameter_dist}. We assumed the availability of spectroscopic redshifts such that, during inference, the true lens and source redshifts were assumed to be known. 

In principle, all masses in the lens environment and line of sight contribute lensing effects. We approximate the entire set of lensing mass as a single strong PEMD perturber plus external shear and convergence ($\kappa_{\rm ext}$). Effectively the density of a uniform mass sheet at the redshift of the main deflector, $\kappa_{\rm ext}$ affects the observed time delays but cannot be constrained by the image positions and fluxes -- a phenomenon called ``mass sheet degeneracy" (MSD) \citep{falco1985model}. For completeness, it should be noted that there exists a separate aspect of MSD that is internal to the main deflector's mass profile, which can be constrained by kinematic tracers of the gravitational potential \citep{koopmans2004gravitational, saha2006gravitational, schneider2013mass, birrer2016mass, shajib2020strides, birrer2020tdcosmohierarchical}. We do not consider this internal mass sheet in our paper.

Failure to account for $\kappa_{\rm ext}$ can bias the $H_0$ inference. The effect of $\kappa_{\rm ext}$ on $D_{\Delta t}$ and $H_0$ is as follows:
\begin{align} \label{eq:kappa_relation}
    D_{\Delta t} \propto \frac{1}{H_0} \propto \frac{1}{1 - \kappa_{\rm ext}}
\end{align}
In time delay cosmography, $\kappa_{\rm ext}$ is often estimated to a few-percent level using tracers of the large scale structure, such as galaxy number counts \citep{rusu2017h0licow} or weak lensing of distant galaxies by all the mass along the line of sight \citep{tihhonova2018h0licow}. Given our focus on assessing the impact of BNN lens modeling on $H_0$, however, we simply place a prior on $\kappa_{\rm ext}$. The images are generated with $\kappa_{\rm ext}=0$ and we draw a true $\kappa_{\rm ext}$ from:
\begin{align} \label{eq:kappa_transformed_prior}
    \frac{1}{1 - \kappa_{\rm ext}} \sim N(1, 0.025)
\end{align}
which, to first approximation, corresponds to
\begin{align} \label{eq:kappa_prior}
   \kappa_{\rm ext} \sim N(0, 0.025)
\end{align}
and translates to an uncertainty of 2.5\% on $D_{\Delta t}$. During inference, we use the exact input distribution in Equation \ref{eq:kappa_transformed_prior} as the $\kappa_{\rm ext}$ prior. While Equation \ref{eq:kappa_transformed_prior} and Equation \ref{eq:kappa_prior} are similar distributions in $\kappa_{\rm ext}$, we choose the former because it amounts to a Gaussian convolution in the $D_{\Delta t}$ posterior, by the relation in Equation \ref{eq:kappa_relation}, whereas the latter introduces non-Gaussianities in the $D_{\Delta t}$ posterior. Note that this choice is strictly numerical and not motivated by the physics. In Section \ref{sec:results}, we discuss further the impact of non-Gaussianities in the individual $D_{\Delta t}$ posteriors on the combined $H_0$. Centering $\kappa_{\rm ext}$ at zero is also an artificial choice. The mean $\kappa_{\rm ext}$ for real lines of sight likely does not vanish for an ensemble of systems due to selection effects, e.g. lens galaxies tend to lie in groups \citep{blandford2000modeling}, causing a slight preference for systems with overdense lines of sight \citep{collett2016observational}.

To simulate measurements of time delays, we artificially added Gaussian errors of 0.25 day to the true time delays, corresponding to zero bias and the smallest possible random errors under current monitoring strategies \cite{ding2018time}. We assumed such an optimistic scenario in time delay measurements so that the $H_0$ inference precision would be dominated by the capabilities of the BNN lens modeling rather than the time delay measurements. 

The true cosmology was a Flat $\Lambda$CDM cosmology with $H_0 = 70 \textrm{km s}^{-1} \textrm{ Mpc}^{-1}$ and $\Omega_{\rm m}=0.3$. Throughout this study, $\Omega_{\rm m}$ was assumed to be known and fixed so only $H_0$ was inferred.

\begin{table}
\begin{center}
\renewcommand{\arraystretch}{1.2}
\begin{tabular}{ l l}
\hline
{\bf Parameter} & {\bf Distribution} \\
\hline\hline
Lens redshift & $z_{\rm lens} \sim N(0.5, 0.2)$ \\
Source redshift & $z_{\rm src} \sim N(2, 0.4)$ \\ 
 \hline\hline
{\bf Lens galaxy} \\
\hline \hline
{Elliptical power-law mass} \\
{Lens center $(^{\prime \prime})$} & $x_\textrm{lens}, y_\textrm{lens} \sim N(0, 0.07)$ \\
Einstein radius $(^{\prime \prime})$ & $\theta_E \sim N(1.1, 0.1)$ \\ 
Power-law slope & $\gamma_{\rm lens} \sim N(2.0, 0.1)$ \\ 
Axis ratio & $q_{\rm lens} \sim N(0.7, 0.15)$ \\ 
Orientation angle (rad) & $\phi_{\rm lens} \sim U(-\pi/2, \pi/2)$ \\  
\hline 
{Elliptical Sersic light} \\ 
Magnitude & $ m_{\rm lens*} \sim U(19, 17)$ \\
Half-light radius $(^{\prime \prime})$ & $R_{\rm lens*} \sim N(0.8, 0.15)$ \\ 
Sersic index & $n_{\rm lens*} \sim N(3, 0.55) $\\
Axis ratio & $q_{\rm lens*} \sim N(0.85, 0.15)$ \\
Orientation angle (rad) & $\phi_{\rm lens*} \sim  U(-\pi/2, \pi/2)$ \\
\hline \hline
{\bf Environment} \\
\hline \hline
External shear modulus & $ \gamma_{\rm ext} \ \sim U(0, 0.05) $ \\
Orientation angle (rad) & $\phi_{\rm ext} \sim U(-\pi/2, \pi/2) $ \\
\hline
External convergence & $ \kappa_{\rm ext} \ \sim N(0, 0.025) $ \\
\hline \hline
{\bf Host galaxy}\\
\hline \hline
{Elliptical Sersic light} \\ 
{Host center $(^{\prime \prime})$} & $x_\textrm{src}, y_\textrm{src} \sim U(-0.2, 0.2)$ \\
Host magnitude & $ m_{\rm src} \sim U(25, 20)$ \\
Half-light radius $(^{\prime \prime})$ & $R_{\rm src} \sim N(0.35, 0.05)$ \\ 
Sersic index & $n_{\rm src} \sim N(3, 0.5) $\\
Axis ratio & $q_{\rm src} \sim N(0.6, 0.1)$ \\
Orientation angle (rad) & $\phi_{\rm src} \sim  U(-\pi/2, \pi/2)$ \\
\hline \hline
{\bf AGN}\\
\hline \hline
{Point source} \\ 
AGN magnitude & $ m_{\rm AGN} \sim U(22.5, 20) $ \\
\hline
\hline
\end{tabular}
\end{center}
\vspace{1.5pt}
\caption{\label{tab:parameter_dist} Parameter distributions}
The distribution of input parameters in the training, validation, and test data. $N(\mu, \sigma)$ denotes a normal distribution with mean $\mu$ and standard deviation $\sigma$ and $U(a, b)$ denotes a uniform distribution with bounds $a$ and $b$.
\end{table}

\subsection{Automated lens modeling with Bayesian neural networks}
\label{sec:automated_lens_modeling}
What sets our method apart from the H0LiCOW method is that the lens model is estimated by the BNN rather than by forward-modeling the images. As indicated in Figure \ref{fig:pipeline_diagram}, the trained BNN takes a test image and outputs the posterior over the target model parameters. The resulting lens model posterior is propagated into $H_0$ inference. There were 11 target model parameters: the six PEMD parameters, the two external shear parameters, the source position coordinates, and the host galaxy size $R_{\rm src}$.  Though not necessary for time delay cosmography, $R_{\rm src}$ was included in our predictions to allow the BNN to explicitly capture its known degeneracy with $\gamma_{\rm lens}$. 

In Section \ref{sec:posterior_inference}, we review the statistical framework of BNN posterior inference in the context of lens modeling. Section \ref{sec:model_and_optim} briefly describes our choices in designing the network architecture and training. More implementation details are available in the Appendix (Section \ref{sec:implementation_details}). 

\subsubsection{Posterior inference}
\label{sec:posterior_inference}
BNNs represent a family of probabilistic neural networks that extends standard neural networks with posterior inference over the network weights \citep{denker1991transforming}. The uncertainty estimated by BNNs can be decomposed into two types: aleatoric and epistemic. Aleatoric uncertainty exists due to the intrinsic randomness in the underlying process. This type of uncertainty would persist even in the limit of infinite training data, i.e. perfect knowledge of the parameter-to-image mapping, because various combinations of parameters may be capable of explaining a given test image. It encodes the $\gamma_{\rm lens}-R_{\rm src}$ degeneracy, for instance; a thick Einstein ring in a test image may be explained by a shallow lens slope or a bigger source. 

Aleatoric uncertainty is explicitly modeled as the width of the distribution over the target parameters. Improving on the work of \cite{hezaveh2017fast} and \cite{levasseur2017uncertainties}, who had used a Gaussian distribution with a diagonal covariance matrix, we adopt a mixture of two Gaussians (henceforward GMM; Gaussian mixture model), each with a full covariance matrix, as we have done in \cite{wagner2020hierarchical}. Explicitly, for a given test lens, we assumed the following form for the distribution over the target lens and light parameters $\xi_{\rm lens}^\star, \xi_{\rm light}^\star$ given the image $d^\star$ and a set of network weights $W$:
\begin{align} \label{eq:aleatoric_posterior}
& p(\xi_{\rm lens}^\star, \xi_{\rm light}^\star | d^\star, W) = w_1(d^\star, W) \phi(\cdot|\mu_1(d^\star, W), \Sigma_1(d^\star, W)) \nonumber \\
& + (1 - w_1(d^\star, W)) \phi(\cdot|\mu_2(d^\star, W), \Sigma_2(d^\star, W)),
\end{align}
where $\phi(\cdot|\mu, \Sigma)$ denotes the PDF of a $p$-dimensional Gaussian with mean $\mu \in \mathbb{R}^p$ and covariance $\Sigma \in \mathbb{R}^{p \times p}$ and the weight on the first Gaussian $w_1 \in (0, \frac{1}{2}]$. The BNN predicted $\mu_1$, $\Sigma_1$, $\mu_2$, $\Sigma_2$, and $w_1$ so the size of the output dimension was $p_{\rm out} = 2 \times \left[p + \frac{p(p+1)}{2}\right] + 1$ for $p$ target parameters. We had $p = 11$, so $p_{\rm out} = 155$. 

Epistemic uncertainty, on the other hand, originates from limited training data or the choice of an imperfect model. It comes into play when the network attempts to generalize to regions outside the training set. In the context of machine learning, it is often described as a distribution over the network weights, post-training. Each realization of the weights corresponds to an alternative model, so integrating over this learned weight posterior is akin to Bayesian model averaging. Folding in the epistemic uncertainty, we have the full predictive distribution: 
\begin{align} \label{eq:predictive_distribution}
    & p(\xi_{\rm lens}^\star, \xi_{\rm light}^\star | d^\star, \Omega_{\rm int}) \nonumber  \\
    &= \int p(\xi_{\rm lens}^\star, \xi_{\rm light}^\star| d^\star , W) p(W| \Omega_{\rm int}) \ dW
\end{align}
where we have made explicit the dependence on the specific training set by appropriately conditioning on the hyperparameters governing the implicit prior, $\Omega_{\rm int}$. Not modeling the epistemic uncertainty at all amounts to simple conditional density estimation, where the weight posterior $p(W | \Omega_{\rm int})$ is a delta function. In standard neural networks, which only give point estimates for the target parameters, both $p(\xi^\star | d^\star , W)$ and $p(W |\Omega_{\rm int})$ are delta functions, so the predictive distribution in Equation \ref{eq:predictive_distribution} collapses to a delta function. 

Consider the integral in Equation \ref{eq:predictive_distribution}. An exact evaluation of this integral is intractable, as it requires averaging over all the weight configurations allowed by $p(W | \Omega_{\rm int})$. There exist several workarounds, including the K-FAC Laplace approximation \citep{mackay1992bayesian, ritter2018scalable}; Bayes by backprop \citep{blundell2015weight}; stochastic MCMC \citep{welling2011bayesian}; deep ensembles \citep{lakshminarayanan2017simple}; and stochastic weight averaging Gaussian (SWAG) variants \citep{maddox2019simple, wilson2020bayesian}. We opt for Monte Carlo (MC) dropout \citep{gal2016dropout, kendall2017uncertainties}, however, for consistency with \cite{wagner2020hierarchical} and simplicity of implementation. In MC dropout, the weight posterior $p(W |\Omega_{\rm int})$ is replaced with the variational distribution $q_\theta(\hat W |\Omega_{\rm int})$ parameterized by $\theta$: 
\begin{align}
& q_\theta(\hat W |\Omega_{\rm int}) = \prod_{i=1}^L q_\theta(\hat W_i |\Omega_{\rm int}) \nonumber \\
& \hat W_i = W_i \cdot \rm{diag}\left(z_{i, j} \right)_{j=1}^{K_i} \nonumber \\
&z_{i, j} \sim \rm{Bernoulli}(p_i) \nonumber \\
&\theta \equiv \{W_i, p_i \}_{i=1}^L
\end{align}
where $i$ indexes the layer of the $L$-layer network and $j$ the node in a given layer. Here, $K_i$ denotes the number of nodes at layer $i$, such that the weight matrix for layer $i$ is $W_i \in \mathbb{R}^{K_i \times K_{i-1}}$. When $z_{i,j} = 0$, the input node $j$ in layer $i$ is dropped out, i.e. set to zero. This form of the variational distribution arises from a mathematical result that a network with randomly dropped weights is equivalent to a deep Gaussian process \citep{damianou2013deep}; see \cite{gal2016dropout} for the derivation. 

To optimize $\theta$, we minimize the KL divergence between the true weight posterior $p(W | \Omega_{\rm int})$ and the variational approximation $q_\theta(\hat W |\Omega_{\rm int})$. Equivalently, the BNN minimizes the log evidence lower bound (ELBO) over the N examples in the training set $\{d^{(n)}, \xi_{\rm lens}^{(n)}, \xi_{\rm light}^{(n)}\}_{n=1}^N$:
\begin{align} \label{eq:bnn_loss}
    \mathcal{L}(W) &= - \sum_{n=1}^N \int q_\theta(\hat W |\Omega_{\rm int}) \log p(\xi_{\rm lens}^{(n)}, \xi_{\rm light}^{(n)}|d^{(n)},\hat W) d\hat W \nonumber \\
    & + \text{KL}(q_\theta(\hat W |\Omega_{\rm int})||p(\hat W)) 
\end{align}
where $p(W)$ is a prior on the network weights. To evaluate the first term in an unbiased way, we approximate each entry in the sum by MC integration with a single sample $\hat W \sim q_\theta(\hat W |\Omega_{\rm int})$. Then $W$ can be updated via stochastic gradient descent (SGD) with respect to the realized sample. The second KL term is the ``regularization'' term that prevents the weights from deviating too far from our prior. This is intractable in its exact form, but reduces to $L_2$ regularization
\begin{align} \label{eq:reg_term}
    \text{KL}(q_\theta(\hat W_i | \Omega_{\rm int})||p(\hat W_i)) \propto \frac{l^2 (1-p_i)}{2N}||\hat W_i||^2
\end{align}
when we assume a prior that can be factorized into a product of Gaussian priors in each layer. The length scale $l$ is a hyperparameter that determines the width of the prior. Note that the dropout probability $p$ is also a hyperparameter in the formulation introduced here. It is not optimized along with $W$ during training and must be tuned manually as part of the hyperparameter search. We assume the same dropout probability $p_i=p_{\rm drop}$ for every layer $i$. For a given choice of $p_{\rm drop}$, $l$ can be folded into the $L_2$ regularization strength hyperparameter $\lambda = l^2 (1-p)/(2N)$. 

The full predictive posterior with the variational approximation is thus
\begin{align} \label{eq:predictive_distribution_variational}
    p(\xi_{\rm lens}^\star, \xi_{\rm light}^\star | \Omega_{\rm int}) &= \int p(\xi_{\rm lens}^\star, \xi_{\rm light}^\star| d^\star , W) q_\theta(W| \Omega_{\rm int}) \ dW.
\end{align}
In order to propagate this into an MCMC-based $H_0$ inference procedure, we need to be able to evaluate it. We do so via MC integration, i.e. by taking some $S$ number of MC dropout iterates and averaging the resulting aleatoric portions of the posterior. 
\begin{align} \label{eq:mc_iterates}
&\hat W^{(s)} \sim q_\theta(\hat W |\Omega_{\rm int}), \quad s=1, \dots, S \nonumber \\
& p(\xi_{\rm lens}^\star, \xi_{\rm light}^\star  | d^\star, \Omega_\text{int}) \approx \frac{1}{S} \sum_{s=1}^S p(\xi_{\rm lens}^\star, \xi_{\rm light}^\star | d^\star , \hat W^{(s)})
\end{align}
The value of $S$ is determined by a convergence test -- that is, they are increased from base values until the full predictive distribution no longer changes. The resulting approximation to the full predictive distribution is a mixture of $S \times 2$ Gaussians, where the factor of $S$ comes from the epistemic MC dropout iterates and $2$ from the aleatoric double-Gaussian parameterization. In particular, the predictive mean is
\begin{align} \label{eq:predictive_mean}
& \mathbb{E}\left[\xi_{\rm lens}^\star, \xi_{\rm light}^\star | d^\star, \Omega_{\rm int}\right] \approx \frac{1}{S} \sum_{s=1}^S \mathbb{E}\left[ \xi_{\rm lens}^\star, \xi_{\rm light}^\star | d^\star , \hat W^{(s)}\right].
\end{align}


\subsubsection{Network architecture and training} \label{sec:model_and_optim}
The convolutional engine of the BNN had the \texttt{ResNet101} architecture \citep{he2015deep}, modified from the \textsc{Torchvision} implementation \citep{marcel2010torchvision}. The specific network architecture used for this paper is illustrated in Figure \ref{fig:network_architecture}. See the Appendix (Section \ref{sec:resnet}) for more details on our choice of architecture.

\begin{figure*}[!htb]
\includegraphics[width=1\textwidth]{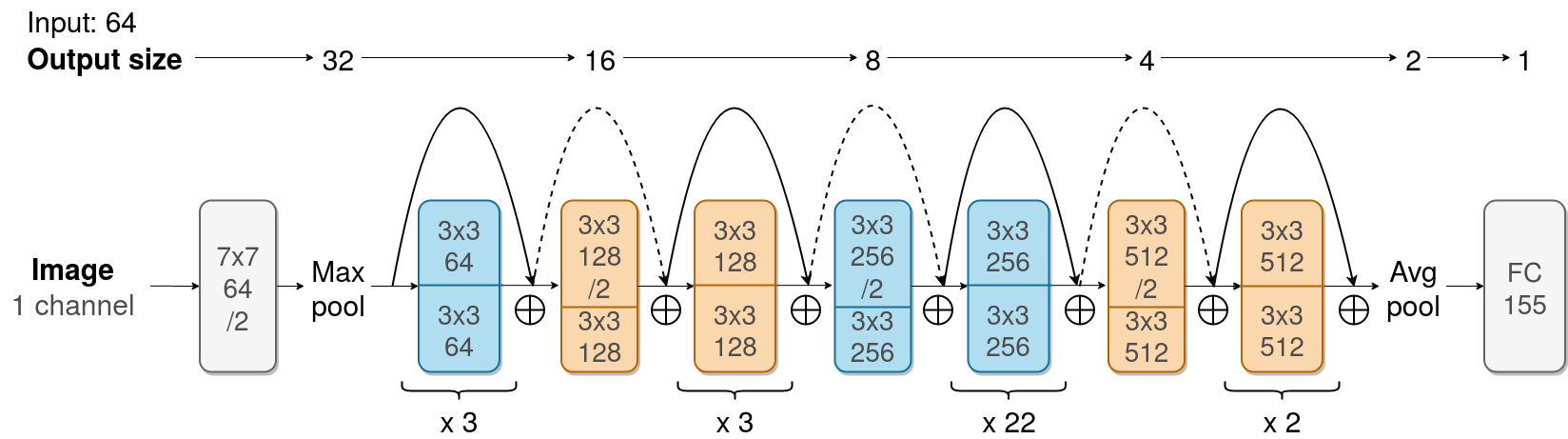}
\caption{The \texttt{ResNet101} network architecture used for the convolutional engine of the BNN. The size of the square feature maps evolves through the layers as indicated on the top. Rectangular boxes contain convolutions of the indicated kernel size and channel number (width). Strides of 2 are denoted as /2. Note that blue and orange boxes are two stacked convolutions. Curved arrows indicate shortcut connections; the solid ones preserve the input feature dimension and dotted ones double the number of channels and halve the feature map resolution. Not shown are the 1D dropout layers, which were inserted before every convolutional layer and before the final fully-connected layer. Batch normalization and ReLU layers followed each convolution as well.}
\label{fig:network_architecture}
\end{figure*}

It is common practice to transform the training images and labels so that they fall into a predefined range. This preprocessing step has the effect of facilitating optimization, as it promotes the numerical stability of the network's hidden units and their gradients. The target labels for the model parameters were normalized so that each parameter had a mean of 0 and standard deviation of 1 across the entire training set. Each input image $d \in \mathbb{R}^{64 \times 64}$ was also pixelwise transformed according to $\log(1 + d_i)$, where $d_i$ represents each pixel intensity value of $d$, and rescaled to the range $[0, 1]$. The log transformation was adopted so that the bright pixels in the cusp of the Sersic or in the AGN images would not overwhelm the informative pixels in the Einstein ring.

The network was trained with the \texttt{ADAM} optimizer \citep{kingma2014adam} for 50 epochs with the weight decay parameter $\lambda = 1\rm{e}{-6}$, batch size $B = 1,024$, and initial learning rate of $\epsilon_0 = 5\rm{e}{-4}$. Although the network was allowed to train for 50 epochs, we only saved the checkpoint with the best validation performance, which plateaued at 37-45 epochs for our experiments. The learning rate was reduced by a factor of 2 whenever the validation loss did not decrease for 50 minibatch updates, until it reached $1\rm{e}{-5}$. The hyperparameters $\lambda$, $B$, and $\epsilon_0$ were tuned via a random search on the validation set. 

\subsubsection{Calibration metric} \label{sec:calibration_metric}
Recall that the MC dropout probability $p_{\rm drop}$ is also a hyperparameter, that is tuned rather than optimized. Among the values $0.5\%$, 0.1\%, and 0\%, the value of 0.1\% was found to be optimal for all three exposure times in our study -- 0.5, 1, and 2 HST orbits. To select a particular dropout probability, we used the confidence-frequency calibration, a semi-quantitative metric introduced in \cite{wagner2020hierarchical}. We reproduce the definition of this calibration metric here and state our own choices in using this metric. 

Denote the $N$ parameter samples drawn from the BNN posterior for some lens $k$ as $\{ \xi_n ^{(k)} \}_{n=1}^{N}$. The true parameter value is $\xi_{\rm true}^{(k)}$. The metric asks: for a given percentage of the BNN posterior probability volume, $p_X$, what percentage of the samples, $p_Y$, contains the truth within this volume? If the posterior is perfectly calibrated, we would expect $p_X$ of the samples to encompass the truth $p_Y = p_X$ of the time, for every value of $p_X$. We can apply this metric on the validation set as a whole by averaging the $p_Y$ values evaluated on individual lenses. To wit,
\begin{align} \label{eq:calibration}
    &p_{ Y}^{(k)}(p_X) = \mathbbm{1}\left\{ \frac{ \sum_{n=1}^N  \mathbbm{1}\left\{d(\xi_n^{(k)}) < d(\xi_{\rm true}^{(k)})\right\}}{N} < p_X \right\} \nonumber \\ 
    &p_{ Y}^{\rm val}(p_X) = \frac{1}{N^{\rm val}} \sum_{k=1}^{N^{\rm val}} p_{ Y}^{(k)}(p_X)
\end{align}
where $\mathbbm{1}\{\cdot \}$ is an indicator function that evaluates to 1 when the argument is true and 0 otherwise, and $d(\xi)$ is a measure of distance of a particular point $\xi$ from the posterior predictive mean given the posterior width. 

Plotting $p_{\rm Y}^{\rm val}$ for a grid of $p_X$ values yields the calibration curve, to be presented and discussed in Section \ref{sec:statistical_consistency} in the context of evaluating the statistical consistency of BNN lens modeling. Regions of the curve with $p_{\rm Y}^{\rm val}<p_X$ speak to an overconfident BNN, because there are not as many lenses with truth within the posterior volume $p_X$ than there should be. Conversely, regions with $p_{\rm Y}^{\rm val}>p_X$ indicate underconfidence.

There are many choices for the distance measure $d$. We use the Mahalanobis distance, a multi-dimensional generalization of the standard score measuring how many standard deviations away a point is from the mean.\footnote{\cite{levasseur2017uncertainties} calibrated their $p_{\rm drop}$ using the 1D standard score. This calibration method was more appropriate for their choice of a diagonal Gaussian as the aleatoric portion of the posterior. For our study using the GMM parameterization, we require the multi-dimensional distance metric so that the parameter covariances can be taken into account.}  
\begin{align} \label{eq:mahalanobis}
    & d(\xi^{(k)}) \equiv \sqrt{(\xi^{(k)} - \mu^{(k)})^T \left[ \Sigma^{(k)} \right]^{-1} (\xi^{(k)} - \mu^{(k)})} \nonumber \\
    & \mu^{(k)} \equiv \mathbb{E}\left[\xi^{(k)} | d^{(k)}, \Omega_{\rm int}\right] \quad \textrm{from Equation \ref{eq:predictive_mean}}  \nonumber \\
    & \Sigma^{(k)} \equiv \rm{Cov}_n \left\{ \xi_n^{(k)} \right\}
\end{align}

\subsection{Individual $H_0$ inference}
\label{sec:individual_h0_inference}
At the stage of inferring $H_0$, we use the lens model posterior obtained from the BNN to properly account for the uncertainties in the lens model parameters. In this section, we describe the inference procedure on the individual lens level: the arrow labeled "2" of the flowchart in Figure \ref{fig:pipeline_diagram}.

We make some simplifying assumptions in our inference. In order to focus on basic $H_0$ recovery, we fix $\Omega_{\rm m} = 0.3$ and infer only $H_0$ for each test lens. In doing so, we only use simulated images and time delay measurements, and do not include velocity dispersion or line of sight measurements in our modeling. Recall also that our training and test data were drawn from the same distribution. In terms of the conventions we introduced in \cite{wagner2020hierarchical}, this setup translates to the assumption that the set of hyperparameters implicit in the training set equals that governing the test prior, i.e. $\Omega_{\rm int} = \Omega$. In addition, we do not place population-level hyperpriors on the individual model parameters; they are assumed to be known and accurate, and thus not varied in a hierarchical manner. This is the approach taken by the {H0LiCOW} collaboration.

The posterior on the test-set hyperparameters $\Omega$ can be written as 
\begin{align}\label{eq:individual_hierarchical_posterior}
    &p\left(\Omega | \Delta t^{(k)}, d ^{(k)} \right) \nonumber \\
     & \propto p(\Omega) p\left(\Delta t^{(k)}, d^{(k)} | \Omega \right) \nonumber \\
    &\propto p(\Omega) \int p\left(\Delta t^{(k)} | \Omega, \xi_{\rm lens}^{(k)}, \kappa_{\rm ext}^{(k)} \right)  \nonumber \\
    & \quad \times p\left(\xi_{\rm lens}^{(k)}, \xi_{\rm light}^{(k)} | d^{(k)}, \Omega_{\rm int}\right) \times \frac{p\left(\xi_{\rm lens}^{(k)}, \xi_{\rm light}^{(k)} | \Omega\right)}{ p\left(\xi_{\rm lens}^{(k)}, \xi_{\rm light}^{(k)} | \Omega_{\rm int} \right)} \nonumber \\
    & \quad \times p\left(\kappa_{\rm ext}^{(k)} \right) 
    \quad d \left( \xi_{\rm lens}^{(k)}\right) d\left(\xi_{\rm light}^{(k)} \right) d \kappa_{\rm ext}^{(k)}
\end{align}
where $k$ denotes a single test lens. The integral in Equation \ref{eq:individual_hierarchical_posterior} represents the total likelihood for this lens. The data are the observed time delay(s) $\Delta t^{(k)}$, where $\Delta t^{(k)} \in \mathbb{R}^1$ for a double and $\Delta t^{(k)} \in \mathbb{R}^3$ for a quad, and the image $d^{(k)}$. The first term in the integral is the time delay likelihood, assumed to be diagonal Gaussian with an uncertainty of 0.25 day. The second-to-last line is the importance-weighted BNN-inferred lens model posterior, which serves as a prior in this level of inference. When the implicit prior differs from the test prior, the BNN posterior $p\left(\xi_{\rm lens}^{(k)}, \xi_{\rm light}^{(k)} | d^{(k)}, \Omega_{\rm int}\right)$ must be divided out by the implicit prior the BNN was trained on and multiplied by the test prior. See \cite{foreman2014exoplanet} and \cite{wagner2020hierarchical} for the derivation of importance-weighting. The external convergence $\kappa_{\rm ext}$ and the lens model parameters $\xi_{\rm lens}^{(k)}, \xi_{\rm light}^{(k)}$ are nuisance hyperparameters that must be integrated out to obtain the population likelihood. 

Applying our assumption of $p(\xi_{\rm lens}^{(k)}, \xi_{\rm light}^{(k)} | \Omega) =p(\xi_{\rm lens}^{(k)}, \xi_{\rm light}^{(k)} | \Omega_{\rm int})$ and paring down the target $\Omega$ to just $H_0$, Equation \ref{eq:individual_hierarchical_posterior} can be greatly simplified to
\begin{align}\label{eq:simplified_individual_hierarchical_posterior}
    &p\left(H_0 |  \Delta t^{(k)}, d^{(k)} \right) \nonumber \\
     & \propto p(H_0) p\left(\Delta t^{(k)}, d^{(k)} | H_0 \right) \nonumber \\
    &\propto p(H_0) \int p\left(\Delta t^{(k)} | D_{\Delta t}^{(k)}(H_0), \xi_{\rm lens}^{(k)},\xi_{\rm light}^{(k)}, \kappa_{\rm ext}^{(k)} \right)  \nonumber \\
    & \quad \times p\left(\xi_{\rm lens}^{(k)}, \xi_{\rm light}^{(k)} | d^{(k)} \right) \nonumber \\
    & \quad \times p\left(\kappa_{\rm ext}^{(k)} \right) 
    \quad d \left( \xi_{\rm lens}^{(k)}, \xi_{\rm light}^{(k)} \right) d \kappa_{\rm ext}^{(k)}.
\end{align}
The notation $D_{\Delta t}(H_0)$ simply makes explicit that, with other cosmological parameters fixed, $D_{\Delta t}$ and $H_0$ have a one-to-one relation for a given lens. 

The individual posterior in Equation \ref{eq:simplified_individual_hierarchical_posterior} is difficult to evaluate due to the complicated dependence structure of $\Delta t^{(k)}$ but lends itself to sampling. We performed MCMC sampling over $D_{\Delta t}$ jointly with $\xi_{\rm lens}^{(k)}, \xi_{\rm light}^{(k)}$, with the following objective evaluated at each MCMC iteration:
\begin{align} \label{eq:mcmc_target}
   &p\left(\Delta t^{(k)} | D_{\Delta t}^{(k)}(H_0), \xi_{\rm lens}^{(k)},\xi_{\rm light}^{(k)}, \kappa_{\rm ext}=0 \right) \nonumber \\
   &\times p\left(\xi_{\rm lens}^{(k)}, \xi_{\rm light}^{(k)} | d^{(k)}\right).
\end{align}
Recall that we do not constrain $\kappa_{\rm ext}$ from data and instead assign a global prior of the form in Equation \ref{eq:kappa_transformed_prior}. To reduce the sampling space, $\kappa_{\rm ext}$ was assumed to be zero during MCMC, and accounted for only in post-processing by multiplying each $D_{\Delta t}$ sample by $\frac{1}{1 - \kappa_{\rm ext}}$ for multiple realizations of $p(\kappa_{\rm ext})$.

The first term in Equation \ref{eq:mcmc_target}, the time delay likelihood, is a 1- or 3-dimensional diagonal Gaussian PDF (depending on whether the lens is a double or a quad). The second term, the BNN-inferred lens model posterior in Equation \ref{eq:predictive_distribution}, does not allow for an exact evaluation but can be approximated using MC integration with $S$ dropout samples, as described in Equation \ref{eq:mc_iterates}. We used $S=12$, making the approximation a mixture of $24$ Gaussians. 

To minimize burn-in time, we initialized the walkers at the positions of the BNN posterior samples along the $\xi_{\rm lens}^{(k)}, \xi_{\rm light}^{(k)}$ dimensions. Along the $D_{\Delta t}$ dimension, the allowed range of the walkers was between 0 and 15,000 Mpc and the initial positions were also uniformly sampled in this range. We found that 18,000 samples gave good coverage of the 12-dimensional sample space (11 for $\xi_{\rm lens}^{(k)}, \xi_{\rm light}^{(k)}$ and 1 for $D_{\Delta t}^{(k)}$). Our implementation uses MCMC sampling modules in \textsc{Lenstronomy}, which uses \textsc{emcee} \citep{foreman2013emcee} to run the sampler. See Table \ref{tab:mcmc_model_params} for a summary of the model parameters and their priors in the cosmological sampling stage. 

Once the $D_{\Delta t}$ MCMC samples were generated this way for each lens, we stored them for the next step of joint-lens inference (Section \ref{sec:joint_h0_inference}). They were, effectively, samples from the individual $D_{\Delta t}$ posteriors $p(D_{\Delta t}^{(k)}| \Delta t^{(k)}, d^{(k)}) \propto p(D_{\Delta t}^{(k)}) p(\Delta t^{(k)}, d^{(k)} | D_{\Delta t}^{(k)})$, when assuming a broad uniform prior $p(D_{\Delta t}^{(k)})$ in the range 0 to 15,000 Mpc.

To obtain the individual $H_0$ posterior, the $D_{\Delta t}$ MCMC samples were converted into $H_0$ using the lens and source redshifts, assumed to be known. Then we applied the uniform $H_0$ prior in the range 50 to 90 $\textrm{km Mpc}^{-1} \textrm{ s}^{-1}$. A Gaussian fit to the resulting $H_0$ samples gave an estimate of the center and spread of $H_0$ posterior for each lens:
\begin{align} \label{eq:H_0_normal}
    H_0^{(k)} \sim N\left(\mu^{(k)}, \sigma^{(k)} \right)
\end{align}

\subsection{Joint $H_0$ inference}
\label{sec:joint_h0_inference}
To perform joint inference on a sample of lenses, we combined the information from the individual $D_{\Delta t}$ posteriors, as indicated in the arrow 5 of the flowchart in Figure \ref{fig:pipeline_diagram}. The $H_0$ posterior from a joint sample is
\begin{align}\label{eq:simplified_joint_hierarchical_posterior}
    &p\left(H_0 | \{ \Delta t\}, \{d\}\right) \nonumber \\
     & \propto p(H_0) p\left(\{\Delta t\}, \{d\} | H_0\right) \nonumber \\
    &\propto p(H_0) \prod_{k} \int p\left(\Delta t^{(k)} | D_{\Delta t}^{(k)}(H_0), \xi_{\rm lens}^{(k)},\xi_{\rm light}^{(k)}, \kappa_{\rm ext}^{(k)}\right)  \nonumber \\
    & \quad \times p\left(\xi_{\rm lens}^{(k)}, \xi_{\rm light}^{(k)} | d^{(k)}\right) \nonumber \\
    & \quad \times p\left(\kappa_{\rm ext}^{(k)}\right) 
    \quad d \left( \xi_{\rm lens}^{(k)}, \xi_{\rm light}^{(k)} \right) d \kappa_{\rm ext}^{(k)}.
\end{align}
where $k$ indexes the test-set lenses. This is identical to Equation \ref{eq:simplified_individual_hierarchical_posterior} in form, except that our data now consist of the entire test set of observed time delays $\{\Delta t^{(k)}\}_{k=1}^{200}$ and images $\{d^{(k)}\}_{k=1}^{200}$. More simply, we can express this joint-sample posterior as
\begin{align}
    p\left(H_0 | \{ \Delta t\}, \{d\}\right) & \propto p(H_0) \prod_k p\left(\Delta t^{(k)} , d^{(k)} | D_{\Delta t}^{(k)}(H_0)\right).
\end{align}

To generate $H_0$ samples from this joint-sample posterior, we MCMC sampled over $H_0$ by evaluating the following likelihood objective at each MCMC iteration: 
\begin{align} \label{eq:joint_mcmc_objective}
    \prod_k p\left(\Delta t^{(k)} , d^{(k)} | D_{\Delta t}^{(k)}(H_0) \right).
\end{align}
Doing so required likelihoods that could be evaluated. Recall, from Section \ref{sec:individual_h0_inference}, that we stored the MCMC samples from individual $D_{\Delta t}$ posteriors. We had applied a broad, uniform prior on $D_{\Delta t}$, so these samples could be reinterpreted as samples from the likelihoods $p\left(\Delta t^{(k)} , d^{(k)} | D_{\Delta t}^{(k)}(H_0) \right)$. What remained was to fit appropriate distributions on these stored samples, so the likelihood could be evaluated. For our main analysis, we adopted the kernel-density estimate (KDE) using Gaussian kernels for its flexibility. The binning scheme followed Scott's normal reference rule \citep{scott2015multivariate}. 

To assess the effect of the fit distribution on the joint-sample inference of $H_0$, we also experimented with two other, less flexible parameterizations of the $D_{\Delta t}$ likelihood. One was the Gaussian parameterization, by which the stored $D_{\Delta t}$ samples for each lens were interpreted to follow a Gaussian distribution and the two Gaussian parameters were fit. That is, we assumed
\begin{align} \label{eq:D_dt_normal}
    D_{\Delta t}^{(k)} \sim N\left(\mu_{D_{\Delta t}}^{(k)}, \sigma_{D_{\Delta t}}^{(k)} \right)
\end{align}
The other was the lognormal parameterization:
\begin{align} \label{eq:D_dt_lognormal}
    \log D_{\Delta t}^{(k)} \sim N\left(m_{D_{\Delta t}}^{(k)}, s_{D_{\Delta t}}^{(k)}\right)
\end{align}

Post-MCMC, we applied our uniform $H_0$ prior to obtain our final, combined $H_0$ posterior. Our implementation of the joint-sample MCMC sampling heavily borrows from \textsc{HierArc}\footnote{\faicon{github} \url{https://github.com/sibirrer/hierArc}} \citep{birrer2020tdcosmohierarchical}, which uses \textsc{Astropy} \citep{robitaille2013astropy} to compute cosmological quantities and \textsc{emcee} to run the sampler.

\begin{table*} 
\centering
\begin{tabular}{l|cc}
    \toprule
    {Parameter} & {Prior} & {Description}  \\
      \midrule
      \textbf{Flat $\Lambda$CDM cosmology} & &  \\
   {$H_0$ ($\textrm{km s}^{-1} \textrm{ Mpc}^{-1}$)} & {$U(50, 90)$} & {Hubble constant}  \\
   {$\Omega_{\rm m}$} & {$\delta(0.3)$} & {Mass density}\\
     \midrule
     \textbf{Mass profile} & &  \\
     {$ \xi_{\rm lens}^{(k)}, \xi_{\rm light}^{(k)}$} & {BNN-inferred lens model posterior } & {PEMD, external shear,} \\
     {} & {(see Equation \ref{eq:mc_iterates})} & {source position/size} \\
     \midrule
     \textbf{Line of sight} & & \\
  {$\kappa_{\rm ext}$} & {$\frac{1}{1 - \kappa_{\rm ext}} \sim N(1, 0.025)$} & {External convergence} \\
    \bottomrule
  \end{tabular}
  \caption{Summary of model parameters}
  \label{tab:mcmc_model_params}
\end{table*}%

\section{Results}
\label{sec:results}
We organize our results as follows. Before we refer to the $H_0$ inference results, in Section \ref{sec:individual_parameter_recovery}, we assess the precision, accuracy, and statistical consistency of the first step in our pipeline: the BNN lens modeling. Having established that individual BNN-inferred lens model posteriors are reasonable, in Section \ref{sec:individual_h0_recovery}, we proceed to interpret the $H_0$ estimates obtained from individual lenses in the context of the TDLMC metrics. In Section \ref{sec:combined_h0_recovery}, we report on the combined $H_0$ predictions and discuss the potential challenges associated with combining information from hundreds of lenses. Section \ref{sec:computational_efficiency} describes the computational efficiency of our pipeline as compared with traditional forward modeling approaches, with cost projections for possible future applications.

See Table \ref{tab:experiments} for a summary of our experiments. The first block of experiments, labeled \textbf{A}, tests the sensitivity of our method to the pixel noise level. We retrain the BNN on images rendered with three different exposure times of 2,700s (0.5 HST orbit), 5,400s (1 HST orbit), and 10,800s (2 HST orbits) for these experiments. The second block \textbf{B} takes either the doubles or the quads from the run with the longest exposure time of 2 HST orbits. There were 89 quads and 111 doubles in our test set of 200 lenses. To control for the sample size, we took all the 89 quads and randomly sampled 89 doubles.

\begin{table*} 
\begin{center}
\begin{tabular}{l|cccccc}
    \toprule
    {Label} & {Exposure time} & {Image} & {Number of} & {Number of} & {Training } & {Inference} \\
    {} & {(HST orbit)} & {configuration} & {test lenses} & {validation lenses} & {time (hr)} & {time (min/lens)}\\
      \midrule
   A1 & {0.5} & {Both}  & {200} & {512} & 5 & 6 \\
   A2 & {1} & {Both}  & {200} & {512} & 5 & 6 \\
   A3 & {2} & {Both}& {200} & {512} & 5 & 6 \\
     \midrule
  B1 &  {1} & {Doubles only}  & {89} & {222} & {-} & 11 \\
  B2 & {1} & {Quads only}  & {89} & {222} & {-} & 6 \\
    \bottomrule
  \end{tabular}
\end{center}
\caption{Summary of experiments}
  Summary of experiments defined by lenses with varying exposure times (block  \textbf{A}) and image configurations (block \textbf{B}). We report the median inference time across the sample of lenses. \label{tab:experiments}
\end{table*}%

\subsection{Individual parameter recovery}
\label{sec:individual_parameter_recovery}

\subsubsection{Accuracy}
To evaluate the accuracy of BNN parameter recovery, we adopt the median absolute error (MAE) metric, defined as the median of the absolute-valued difference between the predictive mean (Equation \ref{eq:predictive_mean}) and the true parameter value across all the lenses in the experiment group, i.e.
\begin{align} \label{eq:mae}
    \textrm{MAE} \equiv \textrm{median}\left\{ \left| \mathbb{E}\left[\xi^{(k)} | d^{(k)}, \Omega_{\rm int}\right] - \xi^{(k)}_{\rm true} \right| \right\}
\end{align}
for each parameter $\xi^{(k)} \in \mathbb{R}$ for lens $k$. Table \ref{tab:mae} lists the MAE values evaluated on the validation set. 

Overall, the BNN yields accurate posteriors. In particular, we can retrieve $\gamma_{\rm lens}$ to 3\% accuracy. Surprisingly, the accuracy does not seem to vary across the exposure times. We investigate the apparent insensitivity to the exposure time in Section \ref{sec:bnn_information}. The 6-7 mas accuracy in the source position is contextualized further in Section \ref{sec:astrometric_requirements}. 

\begin{table*}[]
    \begin{center}
    \renewcommand{\arraystretch}{1.3}
    \begin{tabularx}{\textwidth}{@{} l X X X X X X X X X X X @{}}
    \toprule
\textbf{Experiment} & \textbf{$\gamma_1$} & \textbf{$\gamma_2$} & \textbf{$x_\text{lens} ('')$} & \textbf{$y_\text{lens} ('')$} & \textbf{$e_1$} & \textbf{$e_2$} & \textbf{$\gamma_\text{lens}$} & \textbf{$\theta_E ('')$} & \textbf{$x_\text{src} ('')$} & \textbf{$y_\text{src} ('')$} & \textbf{$R_{\rm src} ('')$}\\ \midrule
0.5 HST orbit  & 0.012 & 0.013 & 0.002 & 0.001 & 0.024 & 0.025 & 0.056 & 0.006 & 0.006 & 0.007 & 0.03\\
1 HST orbit  & 0.012 & 0.013 & 0.002 & 0.002 & 0.025 & 0.025 & 0.056 & 0.006 & 0.007 & 0.006 & 0.03\\
2 HST orbits  & 0.012 & 0.013 & 0.002 & 0.002 & 0.023 & 0.024 & 0.055 & 0.006 & 0.006 & 0.007 & 0.03\\
\midrule
Doubles  & 0.011 & 0.015 & 0.002 & 0.001 & 0.022 & 0.025 & \textbf{0.064} & 0.006 & \textbf{0.008} & \textbf{0.010} & 0.03 \\
Quads  & 0.013 & 0.013 & 0.002 & 0.002 & 0.025 & 0.024 & \textbf{0.050} & 0.006 & \textbf{0.005} & \textbf{0.005} & 0.03 \\
\bottomrule
    \end{tabularx}
    \end{center}
    \caption{Prediction accuracy of individual parameters} 
    Reported values are the median absolute error (MAE) on the validation set (Equation \ref{eq:mae}). Significant differences within each experiment group are bolded.
    \label{tab:mae}
\end{table*}

\subsubsection{Statistical consistency} \label{sec:statistical_consistency}
To probe the statistical consistency of BNN-inferred lens model posteriors with the truth values, we use the calibration metric presented in Section \ref{sec:calibration_metric}.  It is this metric that we used to select our final MC dropout rate $p_{\rm drop}=0.1\%$ as well, from the values $0\%, 0.1\%$ and $0.5\%$.

\begin{figure}[!htb]
\includegraphics[width=0.45\textwidth]{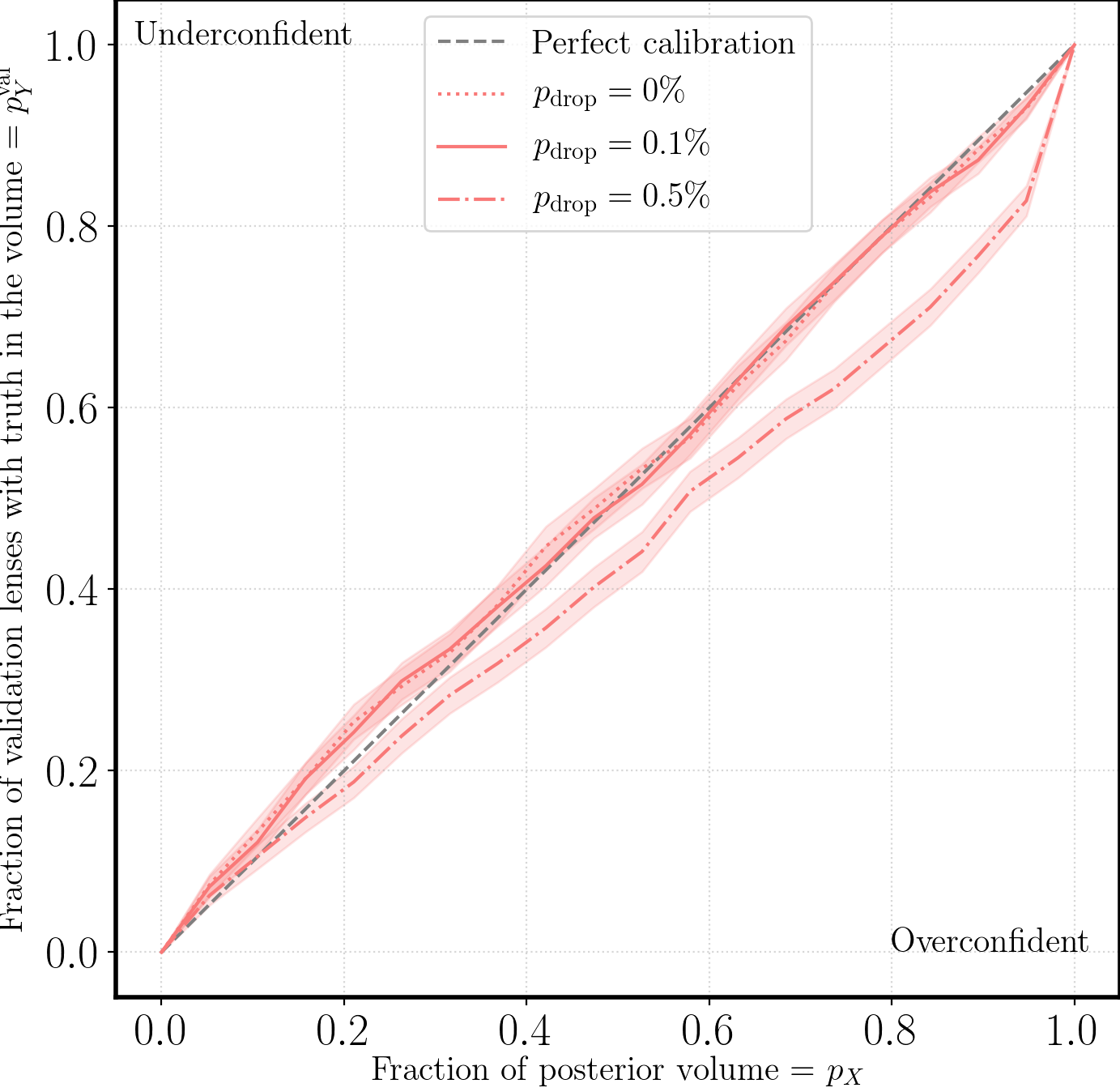}
\caption{The calibration curve $p_Y^{\rm val}$ vs. $p_X$ for the BNN lens model posterior as defined in Equation \ref{eq:calibration}, for an exposure time of 0.5 HST orbit and dropout probabilities of 0\% (no dropout), 0.1\%, and 0.5\%. There were $N^{\rm val} = 512$ lenses in the validation set. The aleatoric portion of the BNN lens model posterior seems to be capturing most of the uncertainty. Not modeling the epistemic uncertainty at all ($p_{\rm drop}=0\%$) does not affect the calibration significantly. A slightly higher dropout probability of $p_{\rm drop}=0.5\%$ leads to overconfidence because the model is underfit. We choose $p_{\rm drop}=0.1\%$.} 
\label{fig:calibration_curve}
\end{figure}

The calibration curves for the exposure time of 0.5 HST orbit and dropout probabilities of 0\% (no dropout), 0.1\%, and 0.5\% are shown in Figure \ref{fig:calibration_curve}. The curves for longer exposure times looked qualitatively similar. The no-dropout and 0.1\% dropout curves are almost indistinguishable; not modeling the epistemic uncertainty at all (i.e. only performing simple conditional density estimation) does not make the model significantly more confident. The epistemic uncertainties are small, most likely because we have a large training set of 512,000 lenses and a relatively flexible GMM parameterization for the aleatoric uncertainty. As mentioned in Section \ref{sec:methods}, we opt for $p_{\rm drop}=0.1\%$ in this paper for all exposure times. We keep the MC dropout parameterization in case the MC dropout captures higher-order epistemic uncertainties that the calibration metric would miss. 

If we increase dropout to $p_{\rm drop}=0.5\%$, the BNN posterior becomes overconfident. This result may be counterintuitive at first glance; a higher dropout probability corresponds to a higher assigned epistemic uncertainty, so we would expect the BNN posterior to become less confident. One possible explanation is the bias-variance tradeoff. Higher dropout means more regularization (lower variance), but regularization can hamper optimization and lead to a larger training error (higher bias). The MAE values for the $p_{\rm drop}=0.5\%$ model were, in fact, higher than those for the $p_{\rm drop}=0.1\%$ model by 30\%. To some extent, this pattern speaks to the limitation of MC dropout as a method of quantifying epistemic uncertainties. It would be worthwhile to explore other methods, such as ensemble-based ones, that are not associated with underfitting. 

\subsubsection{Precision}
Having established that the BNN is accurate and well calibrated, we proceed to report its precision. The parameter uncertainty values in Table \ref{tab:parameter_precision} are the predictive standard deviation, obtained by taking the standard deviation of the posterior samples for each parameter. Similarly as with the MAEs, we report the median uncertainty over the validation set. More concretely, we defined the parameter uncertainty as
\begin{align} \label{eq:parameter_uncertainty}
    & \xi^{(i)(k)} \sim p(\cdot | d^\star, \Omega_\text{int}) \nonumber \\
    & \rm{median}_k \left\{\sqrt{\rm{Var}_i \left\{ \xi^{(i)(k)} \right\} }\right\},
\end{align}
where $\xi^{(i)(k)} \in \mathbb{R}$ refers to the $i^{\rm th}$ posterior sample of this parameter for each lens $k$.

\begin{table*}[]
    \centering
    \renewcommand{\arraystretch}{1.3}
    \begin{tabularx}{\textwidth}{@{} l X X X X X X X X X X X @{}}
    \toprule
\textbf{Experiment} & \textbf{$\gamma_1$} & \textbf{$\gamma_2$} & \textbf{$x_\text{lens} ('')$} & \textbf{$y_\text{lens} ('')$} & \textbf{$e_1$} & \textbf{$e_2$} & \textbf{$\gamma_\text{lens}$} & \textbf{$\theta_E ('')$} & \textbf{$x_\text{src} ('')$} & \textbf{$y_\text{src} ('')$} & \textbf{$R_{\rm src} ('')$}\\ \midrule
0.5 HST orbit  & 0.020 & 0.020 & 0.004 & 0.004 & 0.039 & 0.039 & 0.080 & 0.011 & 0.011 & 0.012 & 0.04\\
1 HST orbit  & 0.020 & 0.020 & 0.005 & 0.005 & 0.039 & 0.040 & 0.077 & 0.011 & 0.012 & 0.012 & 0.04\\
2 HST orbits  & 0.020 & 0.020 & 0.05 & 0.005 & 0.039 & 0.039 & 0.076 & 0.011 & 0.012 & 0.012 & 0.04\\
\midrule
Doubles  & 0.020 & 0.020 & 0.005 & 0.005 & \textbf{0.036} & \textbf{0.036} & 0.074 & 0.011 & \textbf{0.013} & \textbf{0.014} & 0.04 \\
Quads  & 0.020 & 0.020 & 0.005 & 0.005 & \textbf{0.044} & \textbf{0.044} & 0.078 & 0.011 & \textbf{0.011} & \textbf{0.010} & 0.04 \\
\bottomrule
    \end{tabularx}
    \caption{Uncertainty assigned by the BNN on individual parameters}
    Definition is given in Equation \ref{eq:parameter_uncertainty}. Significant differences within each experiment group are bolded.
    \label{tab:parameter_precision}
\end{table*}

Taking the MAE and uncertainty together, the predictions for $\gamma_{\rm lens}$ and $x_{\rm src}, y_{\rm src}$ seem to be more accurate and precise for the quads compared to the doubles. For $\gamma_{\rm lens}$, the accuracy is better by 30\% and precision by 20\%. The source position errors are smaller by 3-5 mas and uncertainty smaller by 2-4 mas. The two extra AGN images in the quads likely offers additional information about the position and orientation of the Einstein ring. Whereas the accuracy on the lens ellipticity parameters $e_1, e_2$ is similar between the doubles and quads, the BNN is 20\% more uncertain about these parameters for the quads on average. The reason for this is not clear and deserves further exploration.

\subsection{Individual $H_0$ recovery}
\label{sec:individual_h0_recovery}
How does the BNN lens modeling performance, validated previously in Section \ref{sec:individual_parameter_recovery}, translate to $H_0$ ($D_{\Delta t}$) for individual lenses? The BNN was sufficiently accurate that, at the level of inferring $H_0$, none of the 200 test lenses were discarded. In this section, we visualize the $H_0$ posteriors for a few lenses and summarize the per-lens $H_0$ recovery for the entire test set of 200 lenses. 

To guide our interpretation of the individual $H_0$ posteriors, let us first introduce a useful benchmark. Aside from the BNN-inferred lens model, two more ingredients affect our $H_0$ inference for a given lens: the time delays and the external convergence. We had assumed small time delay uncertainties and measurement errors (both 0.25 day) so that the relative Fermat potential from the BNN lens modeling would dominate the $H_0$ uncertainty budget (see Equation \ref{eq:relative_time_delays}). We also assumed a narrow distribution in the environment mass densities that would shift the inferred $H_0$ at a 2.5\% level. Whereas the effects of time delays and convergence are small on average, it is instructive to completely isolate the effect of BNN lens modeling on the individual lens level. To this end, we define the ``time delay precision ceiling,'' a reference $H_0$ ``posterior'' which fixes the lens model posterior at the delta-function truth:
\begin{align}\label{eq:time_delay_precision_ceiling}
    &p_{\textrm{precision ceiling}}(H_0 | \Delta t^{(k)}, d^{(k)}) \nonumber \\
    &\propto p(H_0) \int p(\Delta t^{(k)} | D_{\Delta t}^{(k)}(H_0), \xi_{\rm lens}^{(k)},\xi_{\rm light}^{(k)}, \kappa_{\rm ext})  \nonumber \\
    & \quad \times \delta(\xi_{\rm lens}^{(k)} - \xi_{\rm lens, true}^{(k)}) \delta(\xi_{\rm light}^{(k)} - \xi_{\rm light, true}^{(k)}) \nonumber \\
    & \quad \times p(\kappa_{\rm ext})
    \quad d \left( \xi_{\rm lens}^{(k)}, \xi_{\rm light}^{(k)} \right) d \kappa_{\rm ext}.
\end{align}
The time delay precision ceiling represents the precision ceiling in the theoretical case of a perfectly known lens model. 

The $D_{\Delta t}$ posterior under the time delay precision ceiling is exactly Gaussian, by design. To see this, note that
\begin{align} \label{eq:D_dt_relation}
    & D_{\Delta t} = \frac{c \Delta t_{\rm true}}{\Delta \phi} \\
    & \Delta t_{\rm true} = \frac{\Delta t_{\rm obs}}{1 - \kappa_{\rm ext}}
\end{align}
where $\Delta \phi$ is the Fermat potential difference between the images under consideration. The likelihood of the observed time delay $\Delta t_{\rm obs}$ is Gaussian and so is the prior on $\frac{1}{1 - \kappa_{\rm ext}}$ (Equation \ref{eq:kappa_transformed_prior}), so when the lens model is fixed at the truth, $D_{\Delta t}$ being modeled is simply a convolution of Gaussians. 

\begin{figure*}[!htb]
\includegraphics[width=0.95\textwidth]{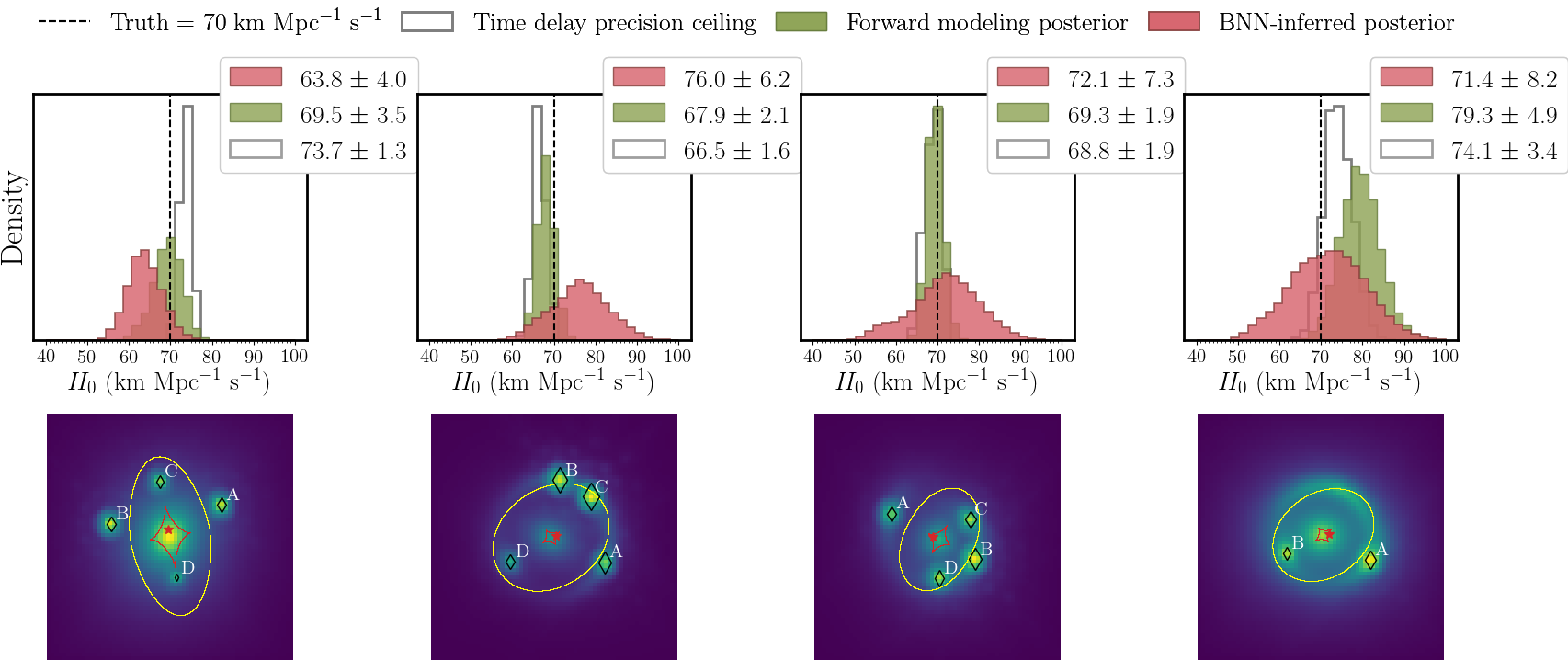}
\caption{Example lenses from the test set with increasing $H_0$ uncertainties from left to right. Top: the BNN-inferred $H_0$ posterior for 2 HST orbits and time delay precision ceiling (defined in Equation \ref{eq:time_delay_precision_ceiling}) for each lens, with the mean and standard deviation (68\% credible interval) of the Gaussian fit (Equation \ref{eq:H_0_normal}). Note that the BNN-inferred posterior and the time delay precision ceiling share the effect of time delay errors and external convergence, so any difference between them is purely due to BNN lens modeling. Bottom: Noiseless images of the lenses, overlaid with the true source position (red star), caustics (red), and critical curves (yellow). AGN image positions are labeled A-D or A-B, with the size of the diamond marker indicating the magnification.}
\label{fig:curated_gallery}
\end{figure*}

Any difference between the BNN-inferred posterior and the time delay precision ceiling can be attributed to the BNN lens modeling. To illustrate this concept on the individual lens level, in Figure \ref{fig:curated_gallery}, we display the BNN-inferred $H_0$ posterior and time delay precision ceiling of four test lenses along with their images. The $H_0$ precision of these four lenses is representative of that in the test set as a whole; the 200 lenses in the test set had been divided into four bins of increasing $H_0$ uncertainty and the four lenses sampled randomly from the bins. Our predictions for the rightmost lens, a double, is the least precise partly because, being spherical, the lens has a very small caustic and hence a short relative time delay. On the other hand, the leftmost, most precise lens is a fold quad.

\begin{figure*}
\includegraphics[width=\textwidth]{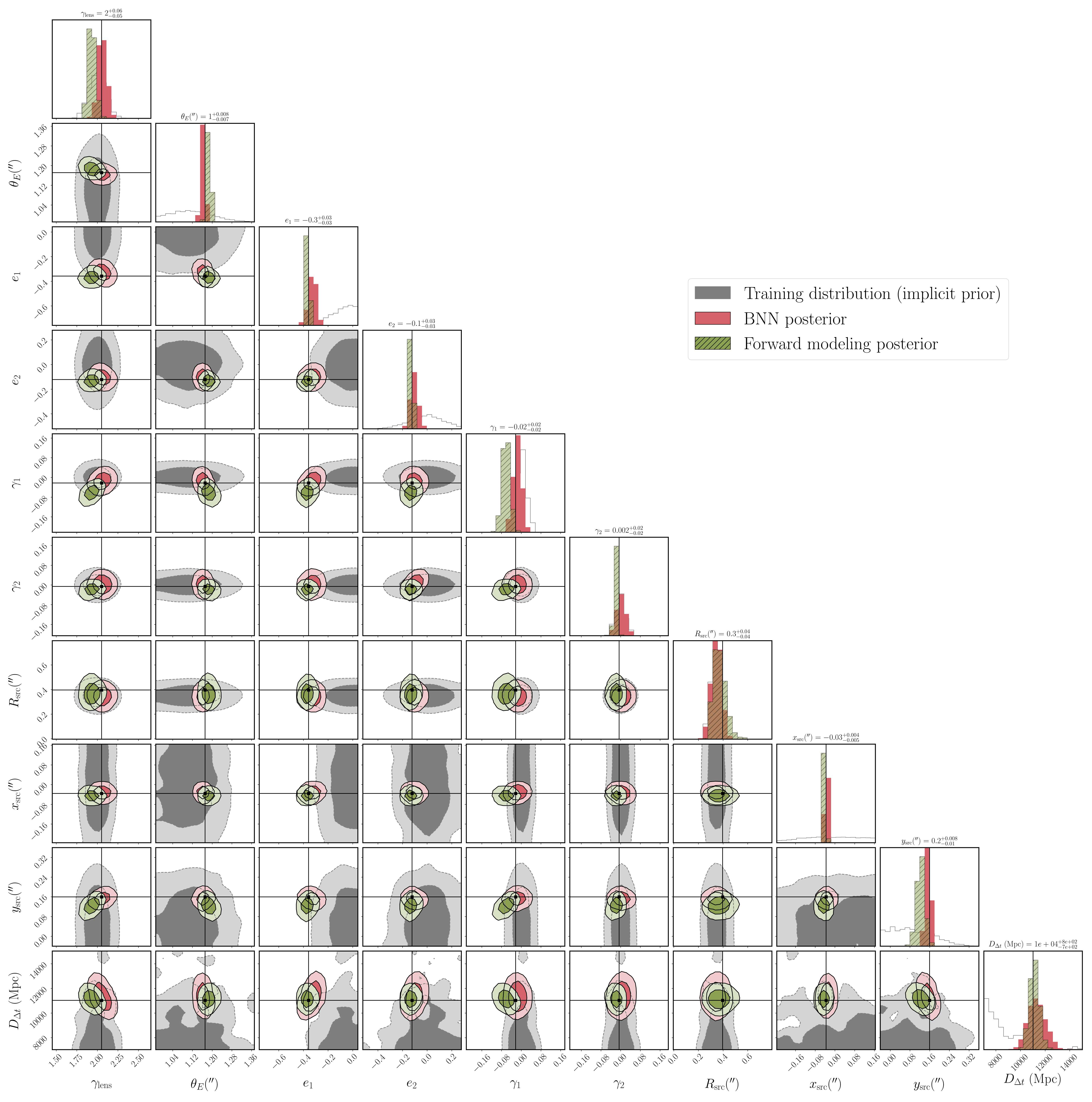}
\caption{The joint posterior over key BNN-predicted parameters and $D_{\Delta t}$ for the leftmost lens in Figure \ref{fig:curated_gallery}. Contours are the 68\% and 95\% credible intervals.}
\label{fig:corner_left}
\end{figure*}

\begin{figure*}
\includegraphics[width=\textwidth]{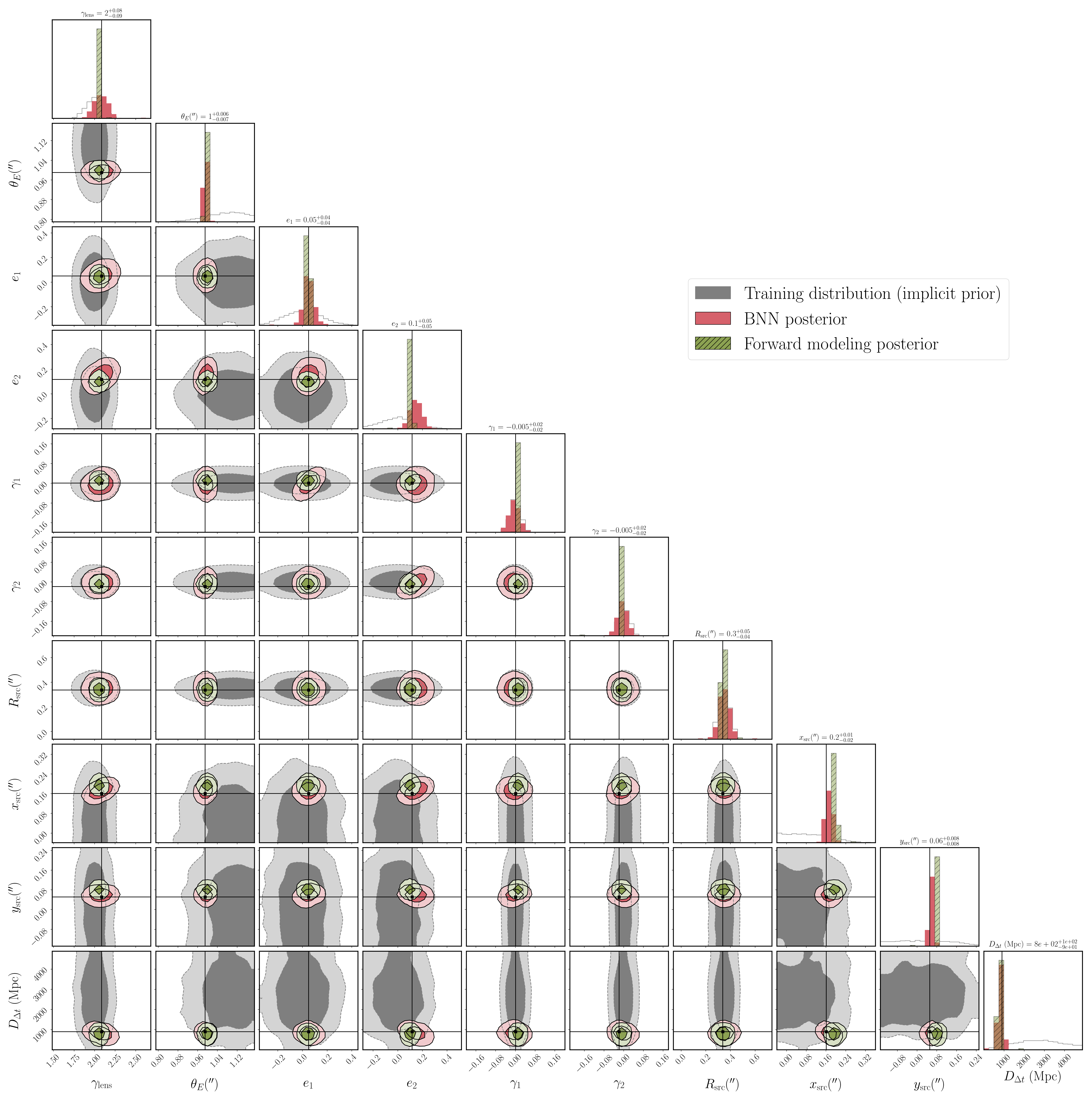}
\caption{The joint posterior over key BNN-predicted parameters and $D_{\Delta t}$ for the rightmost lens in Figure \ref{fig:curated_gallery}. Contours are the 68\% and 95\% credible intervals.}
\label{fig:corner_right}
\end{figure*}

Looking at the joint posteriors over key BNN-predicted parameters and $D_{\Delta t}$ for individual lenses, obtained through the MCMC sampling procedure described in Section \ref{sec:individual_h0_inference}, we can determine whether $D_{\Delta t}$ was sensitive to any particular parameter. Figure \ref{fig:corner_left} and Figure \ref{fig:corner_right} show the posterior over key BNN-predicted parameters and $D_{\Delta t}$ for the leftmost and rightmost lenses in Figure \ref{fig:curated_gallery}. For both lenses, the BNN lens modeling was accurate; the truth falls within the 68\% contour of the inferred posterior. The pairwise correlations between $D_{\Delta t}$ and each parameter reveal that, for the lens in Figure \ref{fig:corner_left}, $D_{\Delta t}$ was mainly sensitive to $\gamma_{\rm lens}$, lens ellipticity, and $y_{\rm src}$. The lens in Figure \ref{fig:corner_right} was sensitive to $\gamma_{\rm lens}$, $e_2$, and $x_{\rm src}$. 

In addition, we can identify pairwise parameter degeneracies modeled by the BNN. Notice that the BNN effectively captures the $e_1-\gamma_1$ and $e_2-\gamma_2$ degeneracies. Because the $R_{\rm src}$ posterior is no better than the implicit prior defined by the training set, however, we do not observe any degeneracy between $\gamma_{\rm lens}$ and $R_{\rm src}$. The BNN does not constrain the external shear parameters $\gamma_1, \gamma_2$ very well beyond the implicit prior, but it assigns sufficiently large uncertainties so as not to bias the $D_{\Delta t}$ inference.

So far, we have looked at the $H_0$ posterior for one lens at a time. To summarize our method's $H_0$ retrieval performance on individual lenses for the test set as a whole, we adopt some of the metrics introduced in the TDLMC: precision, accuracy, and goodness \citep{ding2018time}. Their definitions are reproduced here. Precision ($P$) is the average fractional $H_0$ uncertainty across the $N_{\rm test}$ test lenses in the experimental group. Letting $ \mu^{(k)}, \sigma^{(k)}$ denote the assigned mean and uncertainty of the $H_0$ prediction for lens $k$ from Equation \ref{eq:H_0_normal},
\begin{equation} \label{eq:tdlmc_precision}
    P \equiv \frac{1}{N_{\rm test}} \sum_{k} \frac{\sigma^{(k)}}{H_0}
\end{equation}
Accuracy ($A$) is the average fractional bias across the lenses. If the true $H_0$ value is $H_0$,
\begin{equation} \label{eq:tdlmc_accuracy}
    A \equiv \frac{1}{N_{\rm test}} \sum_{k} \frac{\mu^{(k)} - H_0}{H_0}
\end{equation}
Recall $H_0 = 70 \textrm{ km s}^{-1} \textrm{ Mpc}^{-1}$. Finally, goodness ($\chi^2$) is the standard reduced $\chi^2$ that evaluates the goodness of the assigned uncertainties across the lenses.
\begin{equation} \label{eq:tdlmc_chi2}
    \chi^2 \equiv \frac{1}{N_{\rm test}} \sum_{i} \left( \frac{\mu^{(k)} - H_0}{\sigma^{(k)}} \right)^2
\end{equation}

\begin{figure*}[h]
    \includegraphics[width=0.8\textwidth]{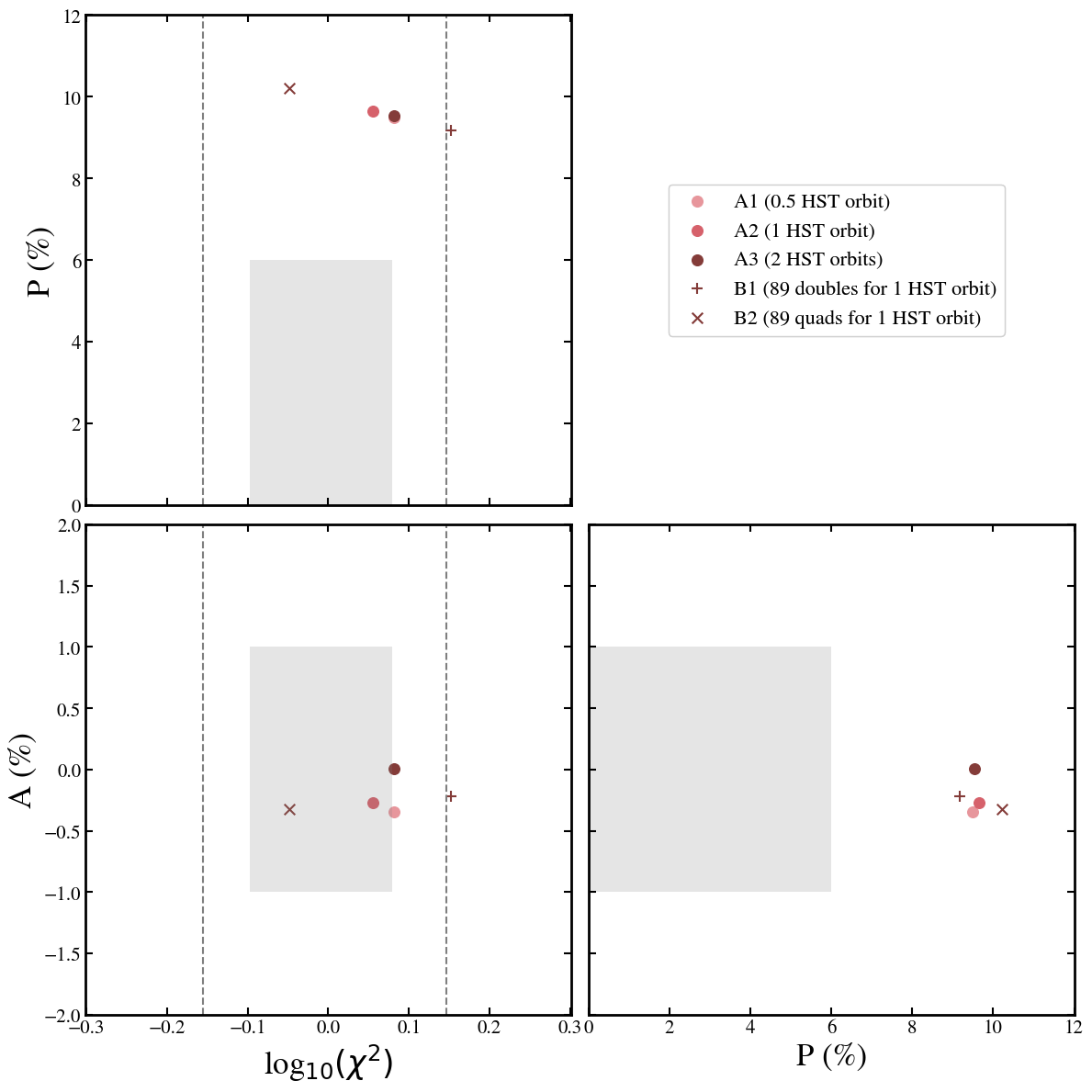}
    \caption{The TDLMC metrics evaluated on the test sample (see Equations \ref{eq:tdlmc_precision} - \ref{eq:tdlmc_chi2} for the definitions). The shaded region corresponds to the target region for a sample of 200 lenses (Equations \ref{eq:tdlmc_precision_target} - \ref{eq:tdlmc_chi2_target}). For the doubles and quads, the dotted lines demarcate the goodness ($\chi^2$) target range for the sample size of 89 (Equation \ref{eq:tdlmc_chi2_doubles_quads_target}). Our precision of $9 -10\%$ lies outside the target of 6\% expected for the best-performing forward modeling approaches. All experiments meet the sub-percent accuracy target. All except the doubles meet the goodness target, but note that the TDLMC metrics weight the lenses equally, so a few outlying lenses could have skewed the metric. \label{fig:tdlmc_corner}}
\end{figure*}

The TDLMC metrics for all our experiments in Table \ref{tab:experiments} are plotted in Figure \ref{fig:tdlmc_corner}. Overlaid are the target ranges for the metrics, which were determined following \cite{ding2018time}. The target for $P$ is based on the best forward-modeling results reported in the literature by the start of the challenge. 
\begin{align} \label{eq:tdlmc_precision_target}
    P < 6\%
\end{align}
Our precision of $9-10\%$ does not meet this target. This is expected, as classical forward modeling would be more precise than the BNN-based inference, even for our simple model assumptions. 

The accuracy target of
\begin{align} \label{eq:tdlmc_accuracy_target}
    |A| < 1\%
\end{align}
expresses the goal of sub-percent accuracy. All the BNN experiments are well within sub-percent accuracy. 

The target range for $\chi^2$ corresponds to the 1 and 99 percentiles of the $\chi^2$ distribution for 200 degrees of freedom (for the 200 lenses in our test set). 
\begin{align} \label{eq:tdlmc_chi2_target}
    0.8 < \chi^2 < 1.2
\end{align}
When taking the 89 doubles or quads only, adjusting the degrees of freedom gives
\begin{align} \label{eq:tdlmc_chi2_doubles_quads_target}
    0.7 < \chi^2 < 1.4.
\end{align}
All except the doubles meet the goodness target, but this does not necessarily imply statistical inconsistency for the doubles, given that the TDLMC metrics weight the lenses equally. In order to account for the varying information content across the lenses, we analyze the combined posteriors in the next section. 

\subsection{Combined $H_0$ recovery}
\label{sec:combined_h0_recovery}
\begin{figure*} 
\includegraphics[width=1\textwidth]{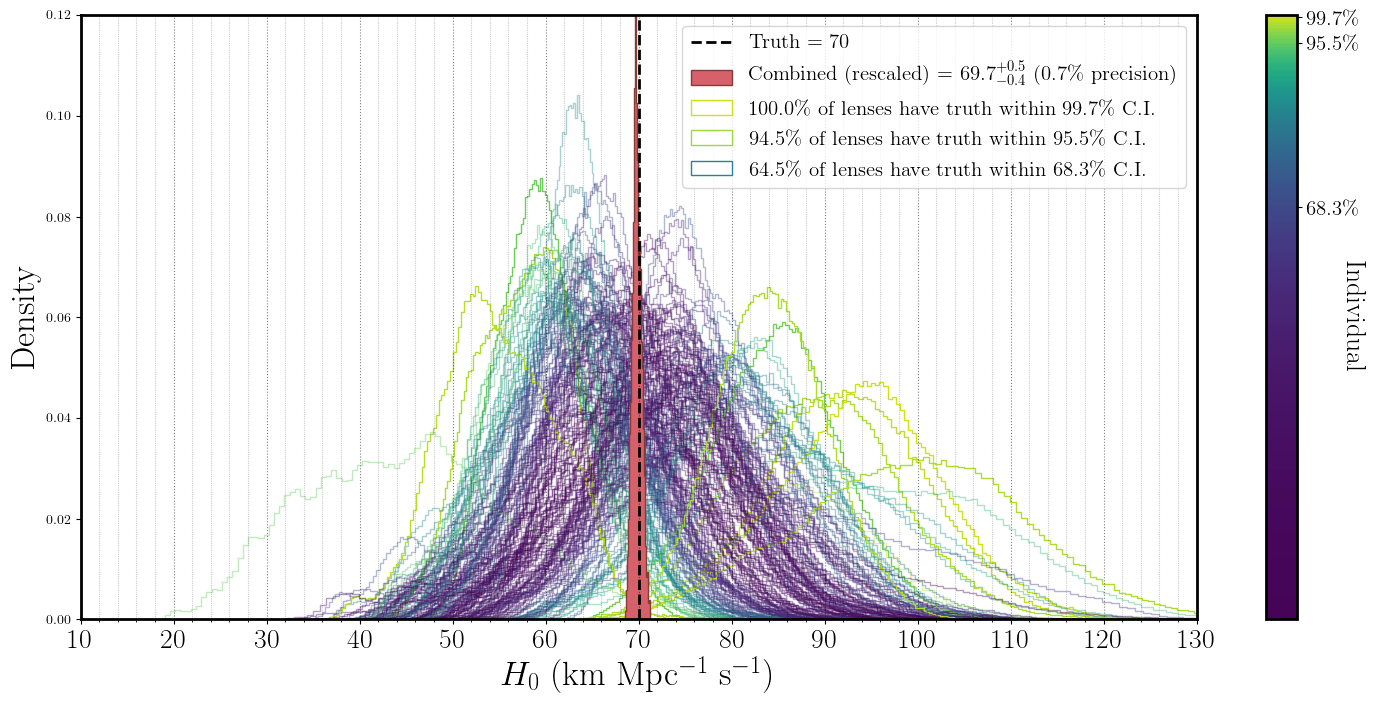}
\caption{The BNN-inferred $H_0$ posteriors for 200 lenses were combined to yield an unbiased $H_0$ estimate of precision 0.5 $\textrm{km s}^{-1} \textrm{ Mpc}^{-1}$ (0.7\%). We overlay the 200 individual $H_0$ posteriors with the joint-sample posterior. The colors of the individual $H_0$ posteriors represent the credible interval in which the truth $H_0$ value falls. The colors corresponding to 68.3\%, 95.5\%, and 99.7\% credible intervals are indicated in the legend. \label{fig:overlaid_200_posteriors}}
\end{figure*}

The true efficacy of our pipeline lies in the joint inference over hundreds of lenses. In Figure \ref{fig:overlaid_200_posteriors}, we overlay the combined $H_0$ posterior for the 200 test lenses along with the 200 individual $H_0$ posteriors, assuming a broad uniform prior in $H_0$ everywhere. We report a final precision of 0.5 $\textrm{km s}^{-1} \textrm{ Mpc}^{-1}$ (0.7\%). There is no detectable bias; the combined posterior is consistent with the truth. As an additional test of statistical consistency, for each lens, we computed the credible interval at which the truth $H_0$ lies. The confidence matched frequency closely; the truth fell within the 68.3\% credible interval in 65\% of the lenses, within the 95.5\% credible interval in 95.5\% of the lenses, and within 99.7\% credible interval in 99.0\% of the lenses. 

\begin{figure} 
\includegraphics[width=0.45\textwidth]{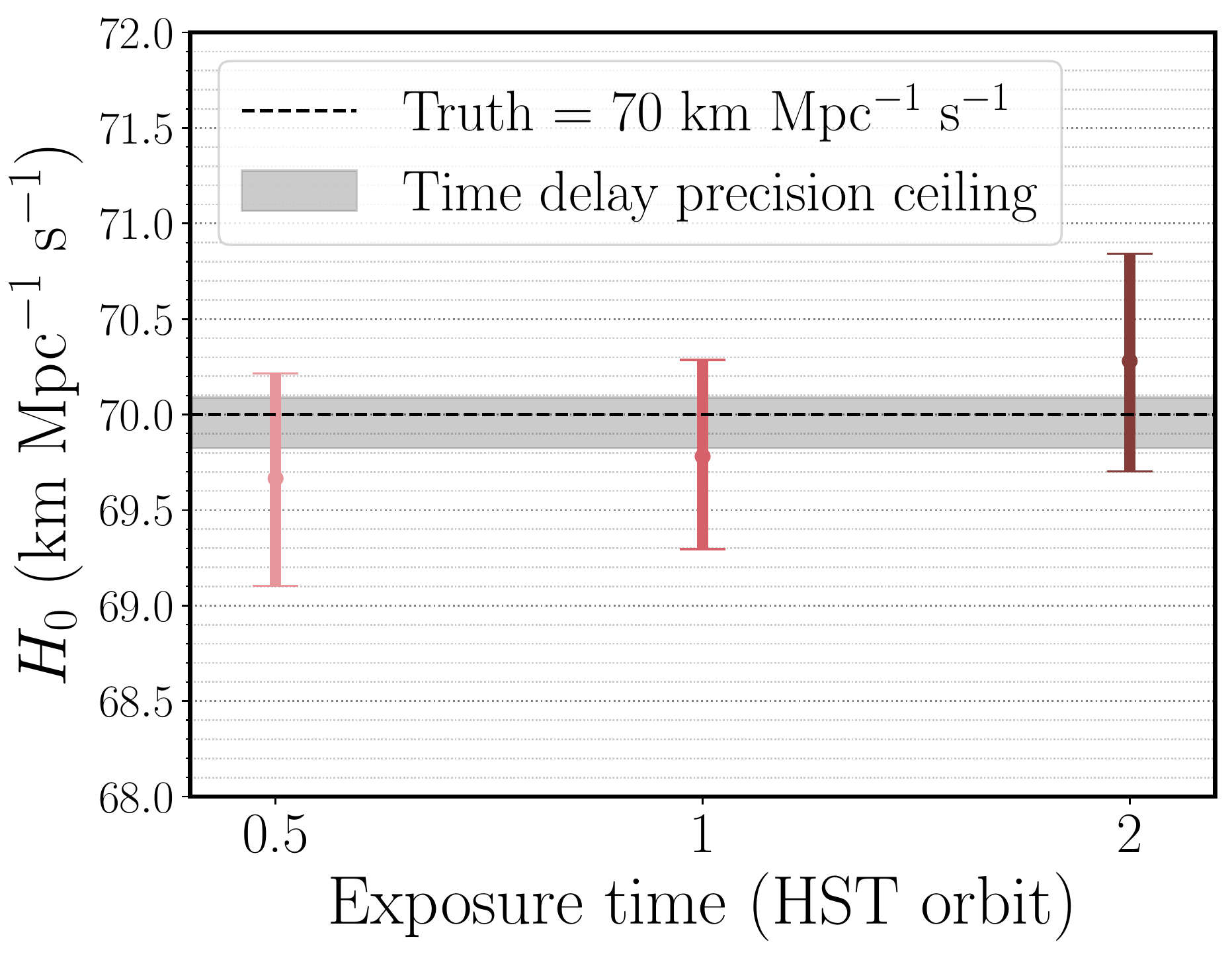}
\caption{Combined $H_0$ prediction on 200 test-set lenses (solid). Error bars represent 68\% credible intervals. The final precision is 0.5 $\textrm{km s}^{-1} \textrm{ Mpc}^{-1}$ (0.7\%) and there is no detectable bias. \label{fig:boxplot_exptime}}
\end{figure}

To assess the impact of the exposure time and image configuration on the joint-sample inference, the combination was done for each of the experiment groups in Table \ref{tab:experiments}. Figure \ref{fig:boxplot_exptime} shows the combined estimate for experiments A1-A3, i.e. 200 lenses in the exposure times of 0.5, 1, and 2 HST orbit(s). The precision of the combined posterior does not vary with image depth, as expected from the lack of such a trend in the individual parameter recovery (Section \ref{sec:individual_parameter_recovery}) and individual $H_0$ recovery (Section \ref{sec:individual_h0_recovery}) with the image depth.

\begin{figure}
\includegraphics[width=0.45\textwidth]{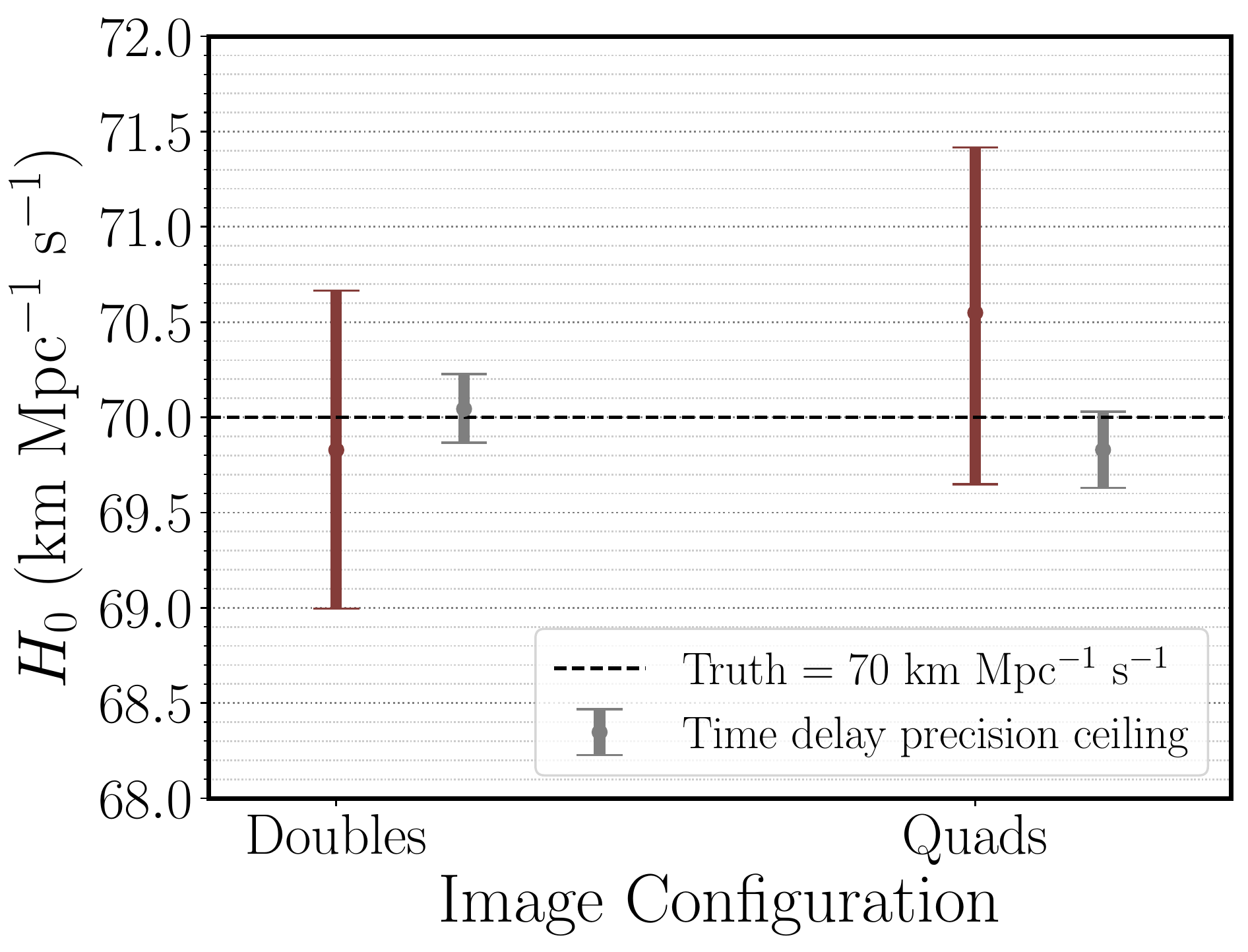}
\caption{Combined $H_0$ prediction on 89 doubles and 89 quads (solid). Error bars represent 68\% credible intervals. For both doubles and quads, the final precision is around $0.7$ $\textrm{km s}^{-1} \textrm{ Mpc}^{-1}$ ($1\%$) and there is no detectable bias. Again, the combined posteriors do not vary significantly with image depth.}
\label{fig:boxplot_doubles_vs_quads}
\end{figure}

Figure \ref{fig:boxplot_doubles_vs_quads} shows the combined $H_0$ posteriors for the 89 doubles and 89 quads separately (groups B1, B2 in Table \ref{tab:experiments}). For both doubles and quads, the precision is comparable, at $0.7$ $\textrm{km s}^{-1} \textrm{ Mpc}^{-1}$ ($1$\%), and there is no bias. Again, there is no significant trend with image depth.

Systematics that appear small on an individual level can figure prominently in the combined posterior when the sample size is large. We find that the form of the fit distribution for the $D_{\Delta t}$ posterior merits careful consideration. See Section \ref{sec:ddt_parameterization} for a discussion of this issue. 

\section{Discussion}

\subsection{Pixel-level information processed by the BNN} \label{sec:bnn_information}
In this section, we explore the patterns in the BNN predictions with the exposure time and various properties of the lens, to confirm that the BNN-inferred posteriors are reasonable. We also compare the BNN lens modeling with traditional forward modeling on a lens-by-lens basis.

\begin{figure}[t]
\includegraphics[width=0.45\textwidth]{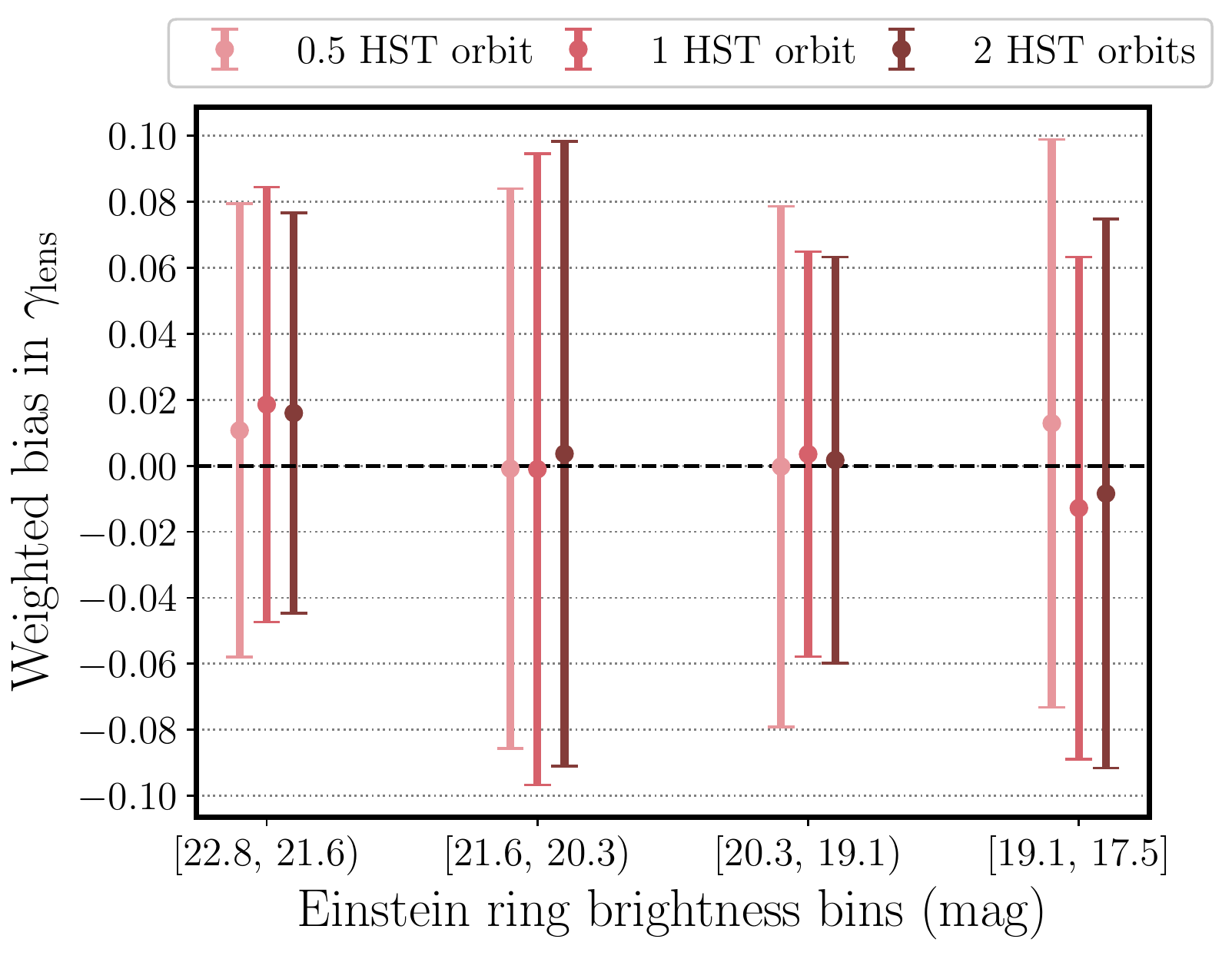}
\caption{The weighted mean and standard deviation of the absolute bias in $\gamma_{\rm lens}$, binned by the Einstein ring brightness. The bias is consistent with zero. There is no strong pattern in the center and spread of the bias with the ring brightness in any of the exposure times.}
\label{fig:gamma_vs_ring_brightness}
\end{figure}

According to the MAE values in Table \ref{tab:mae} and BNN-assigned parameter uncertainty in Table \ref{tab:parameter_precision}, the BNN appears to be insensitive to the exposure time. To investigate this surprising behavior, we take a closer look at the BNN constraints on the $\gamma_{\rm lens}$ parameter, whose information is largely contained in pixels comprising the Einstein ring. In Figure \ref{fig:gamma_vs_ring_brightness}, we plot the weighted mean of the absolute bias in $\gamma_{\rm lens}$ binned by the surface brightness of the Einstein ring, where the weights are the inverse of the predictive variances in $\gamma_{\rm lens}$. The surface brightness of the Einstein ring was computed by rendering an image of the source galaxy, without the lens light, and summing up the pixel values. The error bars indicate the weighted standard deviation of the absolute bias. We confirm that the bias is consistent with zero, so $\gamma_{\rm lens}$ does not bias our $H_0$ predictions downstream. Yet we do not detect a strong correlation in the center and spread of the $\gamma_{\rm lens}$ bias with the ring brightness, nor with the exposure time.

Does the apparent lack of dependence on the exposure time originate from the data or the BNN? To explore this question, we directly compare the BNN to forward modeling. Because running forward modeling on all 200 test lenses is computationally prohibitive, we examine one lens at a time. In Figures \ref{fig:corner_left}, and \ref{fig:corner_right}, we overlay the BNN-inferred posterior with the forward modeling posterior. Forward modeling achieves tighter constraints on most parameters by a factor of 2--3 compared to the BNN. In Figure \ref{fig:curated_gallery}, we see that forward modeling can even approach the time delay precision ceiling on $H_0$ on some lenses whereas the BNN precision trails behind. The BNN is indeed limited in its ability to extract all the information out of the pixels. The strength of the BNN method lies not in the per-lens precision but in the accuracy and computational efficiency, which enable joint inference over hundreds of lenses.

\begin{figure}[t]
\includegraphics[width=0.45\textwidth]{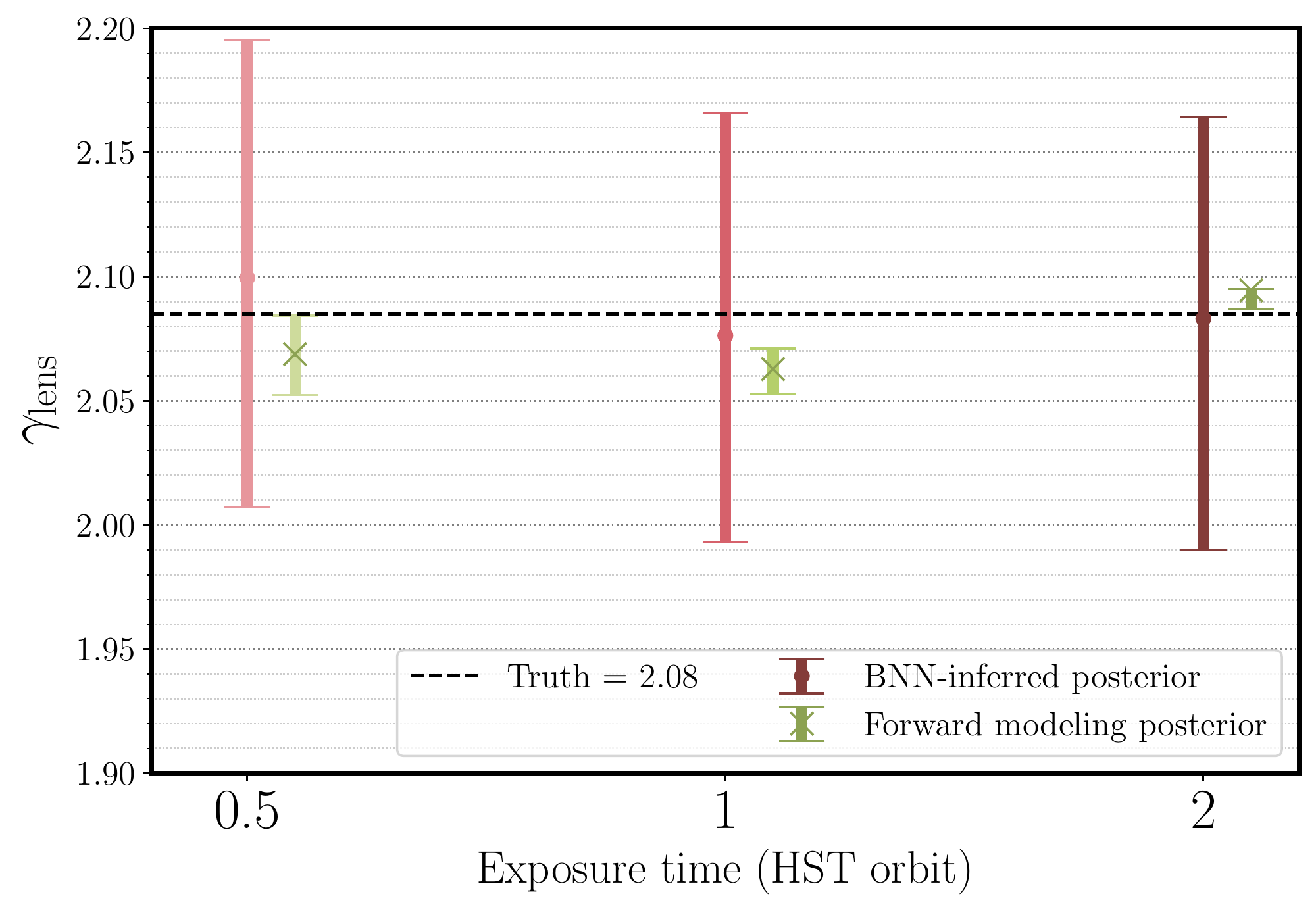}
\caption{The marginal posteriors on $\gamma_{\rm lens}$ from forward modeling and BNN for the rightmost lens in Figure \ref{fig:curated_gallery}. Forward modeling generates tighter constraints on deeper images whereas the BNN precision remains similar.}
\label{fig:fm_exposure_time}
\end{figure}

There is more information in the pixels than the BNN can harness at each exposure time. In addition, with increasing exposure time, the forward modeling constraints become tighter whereas the BNN constraints remain at the same level. Figure \ref{fig:fm_exposure_time} overlays the constraints on $\gamma_{\rm lens}$ from the BNN and forward modeling across the exposure times, for a single lens. Forward modeling is able to extract the extra information from deeper images, achieving higher precision by a factor of 6, 8, and 13 compared to the BNN for exposure times of 0.5, 1, and 2 HST orbits, respectively. The inner workings of the BNN's response to image depth requires further investigation. We postpone this to future work.

\begin{figure}[!htb]
\includegraphics[width=0.45\textwidth]{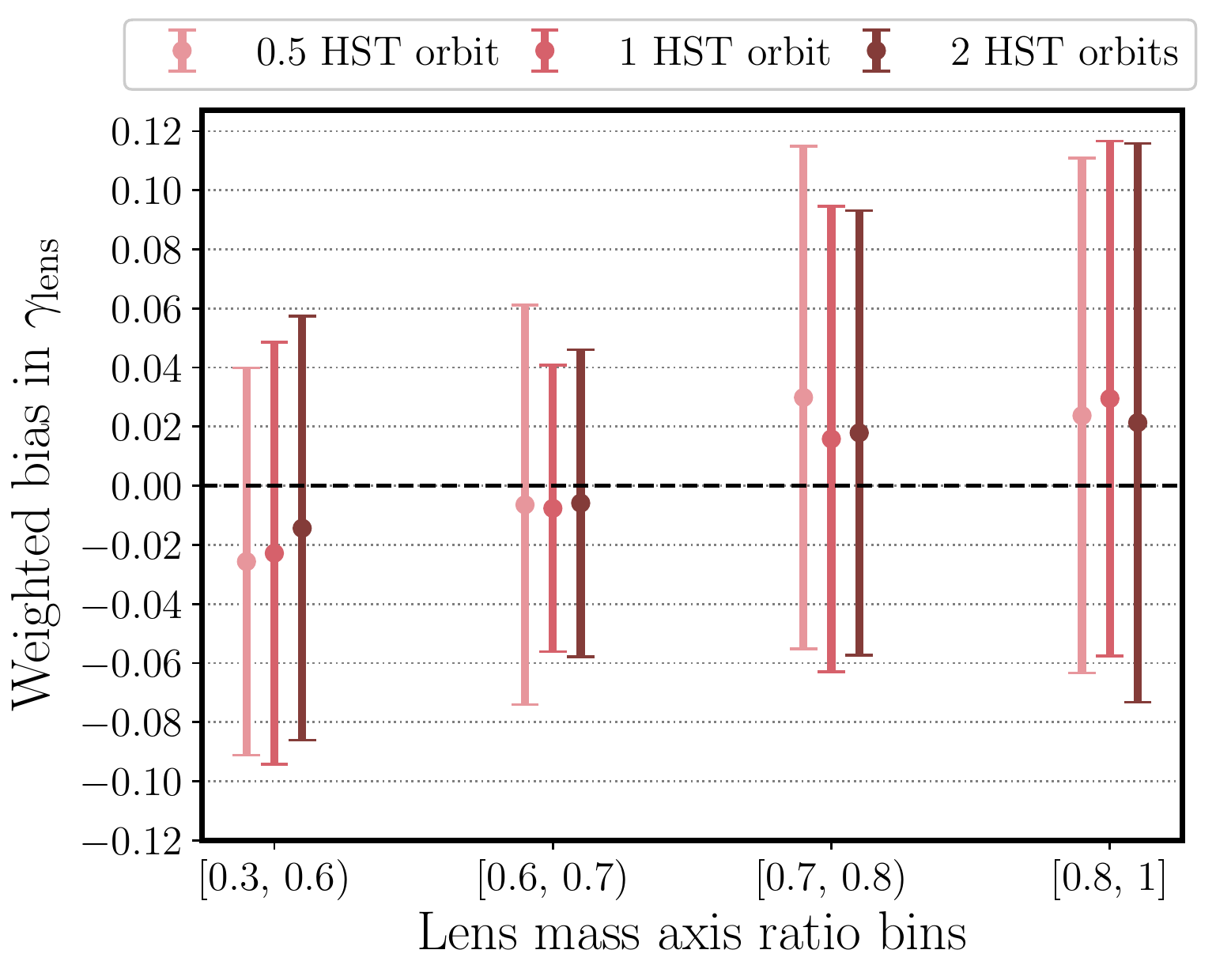}
\caption{The weighted mean and standard deviation of the absolute bias in $\gamma_{\rm lens}$, binned by the Einstein ring brightness. The spread of the bias clearly increases for more spherical lenses.}
\label{fig:gamma_vs_ellipticity}
\end{figure}

\begin{figure}[!htb]
\includegraphics[width=0.45\textwidth]{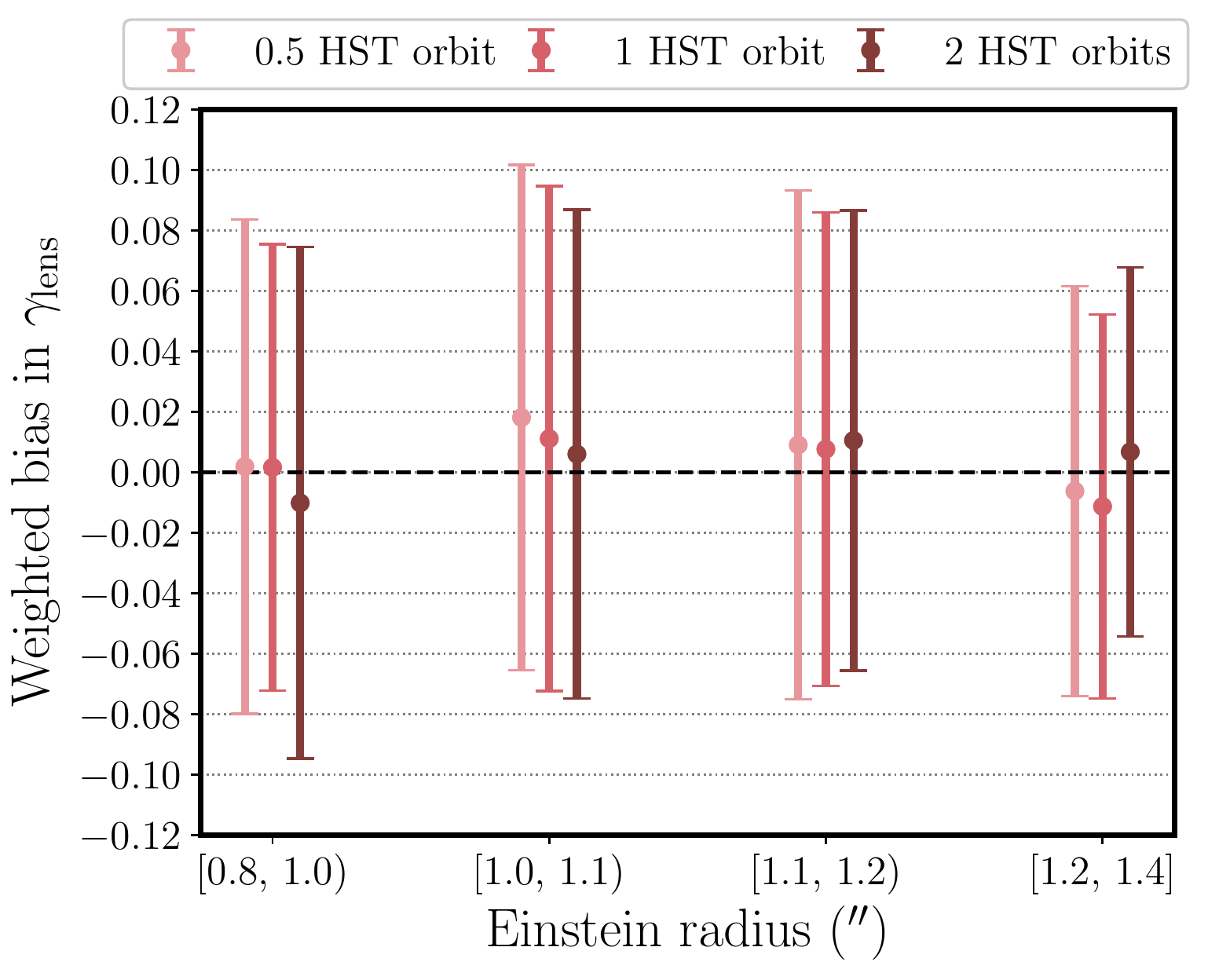}
\caption{The weighted mean and standard deviation of the absolute bias in $\gamma_{\rm lens}$, binned by the Einstein radius. Bigger-separation lenses lead to a small spread in bias, as expected.}
\label{fig:gamma_vs_theta_E}
\end{figure}

Given that the pixel intensity values do not significantly affect the BNN, we proceed to establish whether the BNN predictions follow expected trends with more geometric features of the image. In Figure \ref{fig:gamma_vs_ellipticity}, we plot the weighted mean of the absolute bias in $\gamma_{\rm lens}$ binned by the lens axis ratio, where the weights are the inverse of the predictive variances in $\gamma_{\rm lens}$. We see a trend of smaller uncertainty in $\gamma_{\rm lens}$ with more elliptical lenses, most likely because the elongated shape of the critical curves make the relative Fermat potential differences more dramatic. Because doubles are generally more spherical than quads, this trend may partly explain why the parameter constraints on $\gamma_{\rm lens}$ was 30\% more accurate and 20\% more precise for quads compared to doubles (see Tables \ref{tab:mae} and \ref{tab:parameter_precision}). Similarly, in Figure \ref{fig:gamma_vs_theta_E}, we bin by the image separation and find that the spread of the bias reduces for lenses with bigger separation, as we might expect.

\subsection{Astrometric requirements} \label{sec:astrometric_requirements}
We can put the BNN's source position predictions in the context of time delay cosmography by propagating astrometric errors through the relative Fermat potential, into the $H_0$ inference (see Equation \ref{eq:constraint_D_dt}). \cite{birrer2019astrometric} derive the following approximate requirement for the astrometric uncertainty to be sub-dominant to the time delay uncertainty:
\begin{align} \label{eq:astrometric_18}
    \theta_{i, j} \sigma_\beta \lesssim \sigma_{\Delta t_{i, j}} \frac{c}{D_{\Delta t}}
\end{align}
where $\theta_{i, j}$ is the separation between images $i$ and $j$; $\sigma_{\Delta t_{i, j}}$ the uncertainty in the measurement of the relative time delay, $\Delta t_{i, j}$; and $\sigma_{\beta}$ the astrometric uncertainty in the source plane. For our galaxy-scale lenses, $\theta_{i, j} \sim 1 ''$. Further assuming $z_{\rm lens} \sim 0.5, z_{\rm src} \sim 2$ and $\Delta t_{i, j} \sim 4$ days, we can estimate that the astrometric uncertainty will have to be beaten down to 3 mas. The BNN's astrometric uncertainties were 6-7 mas, as estimated from adding in quadrature the median BNN-assigned uncertainties on $x_{\rm lens}/y_{\rm lens}$ and $x_{\rm src}/y_{\rm src}$ -- because the BNN was trained to predict $x_{\rm src}/y_{\rm src}$, defined as the source's offset from the lens centroid $x_{\rm lens}/y_{\rm lens}$. For the BNN, the astrometric uncertainties will therefore dominate the uncertainty error budget. Conversely, for $\sigma_\beta \sim 7$ mas to be sub-dominant, the time delay measurement would have been degraded to $\sigma_{\Delta t_{i, j}} \sim 0.6$ day.

\subsection{Impact of the lens light in BNN lens modeling}
By comparing these MAE values with the values reported in \cite{wagner2020hierarchical}, where the BNN was trained on lens-subtracted images, we can draw rough conclusions about the impact of including the lens light on parameter recovery. For one, the lens light seems to have aided in the prediction of the lens centroid, because we artificially centered the lens light with the lens mass. Our predictions for $x_{\rm lens}, y_{\rm lens}$ were accurate to 1--3 mas, compared to 5--6 mas on lens-subracted images for the best-performing BNN model in \cite{wagner2020hierarchical}(Gaussian mixture model, 0.1\% dropout). The partial mixing of Einstein ring with the bright lens light, however, appears to have hurt the prediction accuracy for most other parameters, notably $\gamma_{\rm lens}$ which had double the MAE. Other parameters saw an MAE increase of 25-50\%. The interpretation that the lens light hurts more than help agrees with the previous experiments with non-Bayesian convolutional neural nets, in which the lens light reduced parameter accuracy by 30-40\% \citep{pearson2019use}. Note, however, that this comparison is approximate at best, as it does not control for the size of the training set (our 512,000 vs. previous 400,000), the width of the training distribution (previous was almost twice as wide), and the volume of the target parameter space (our 11 dimensions vs. previous 8).

\subsection{Choice of fit distribution} \label{sec:ddt_parameterization}

\begin{figure*} 
\includegraphics[width=0.9\textwidth]{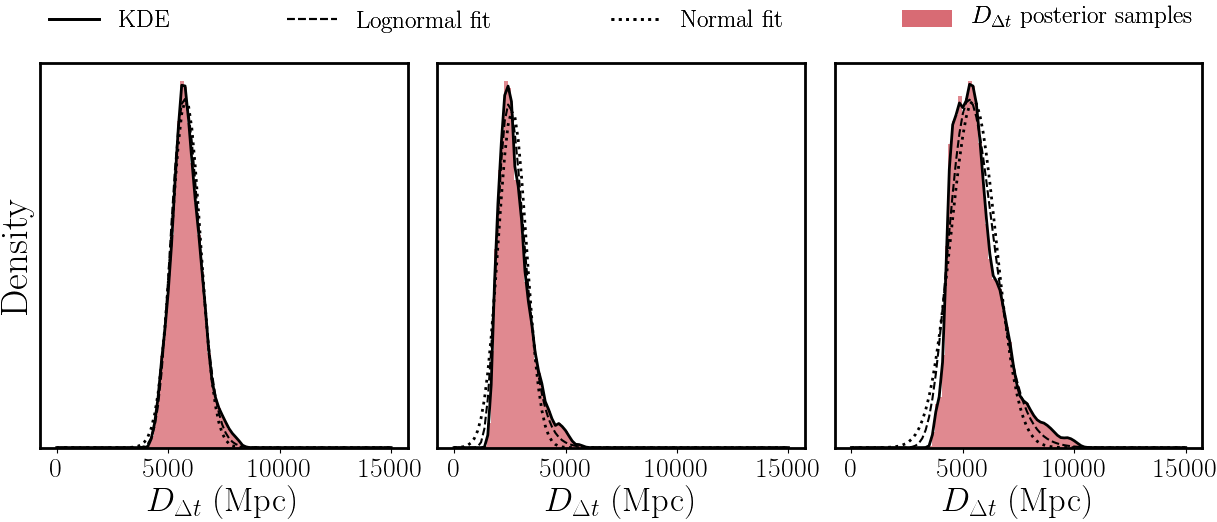}
\caption{A simple illustration of the importance of the fit distribution. When the dominant source of uncertainty in $D_{\Delta t}^{(k)}$ is the relative Fermat potential, the $D_{\Delta t}^{(k)}$ posterior will take on a heavy upper tail. KDE can capture the tail well. On the other hand, the lognormal (Equation \ref{eq:D_dt_lognormal}) tail is not heavy enough for some lenses. Normal approximation (Equation \ref{eq:D_dt_normal}) will always underestimate $D_{\Delta t}^{(k)}$, even compared to lognormal. \label{fig:kde_lognormal_vs_normal}}
\end{figure*}

When the uncertainty in the relative Fermat potential is the precision bottleneck and is approximately Gaussian, it follows from Equation \ref{eq:D_dt_relation} that $D_{\Delta t} \propto  1/{{\rm Gaussian}}$. The PDF for the inverse of a Gaussian random variable does not exist \citep{robert1991generalized} but the resulting distribution has a heavy upper tail. We opted to run KDE on the MCMC samples from the $D_{\Delta t}$ posterior (as stated in Equation \ref{eq:D_dt_lognormal}) so as to explicitly assign weight to this upper tail. Gaussian or lognormal distributions, on the other hand, will not be appropriate. As a simple illustrative example, a Gaussian fit will underestimate $D_{\Delta t}^{(k)}$. See Figure \ref{fig:kde_lognormal_vs_normal} for a visual comparison of lognormal and normal fits on individual $D_{\Delta t}^{(k)}$ samples. The normal fit always lies to the left of the lognormal fit. The difference is qualitatively small for some lenses with a well-constrained Fermat potential, e.g. the leftmost lens, but is consistent. 

\begin{figure} 
\includegraphics[width=0.45\textwidth]{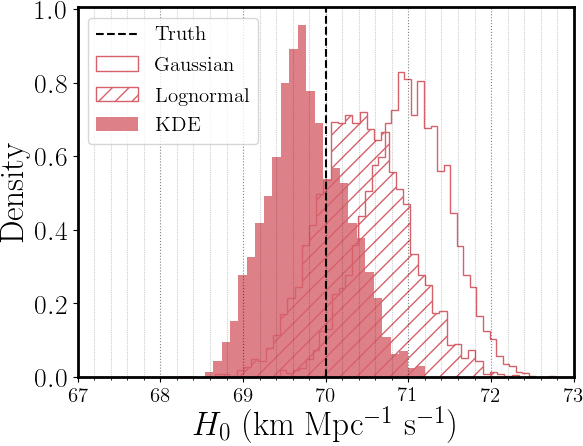}
\caption{Small fit errors on individual $D_{\Delta t}$ posteriors can be amplified into a substantial bias in the combined $H_0$ when the sample size is 200. For a simple illustration, we overlay the combined $H_0$ from using a Gaussian fit distribution -- and observe a significant upward bias. Lognormal does better because it can still capture some of the upper tail in $D_{\Delta t}$. KDE agrees the best with the truth. \label{fig:kde_lognormal_vs_normal_hist}}
\end{figure}

In Figure \ref{fig:kde_lognormal_vs_normal_hist}, we confirm that the Gaussian is an inadequate choice of the fit distribution when the sample size is 200; seemingly small fit errors on individual lenses results in a significant upward bias in the combined $H_0$. Fitting a lognormal instead brings the combined $H_0$ posterior to a level more consistent with the truth. The KDE parameterization agrees the best with the truth.
 
As we look to the prospect of hundreds, maybe thousands, of lens samples, making statements at a $0.1\%$ level will require knowing the shape of the $D_{\Delta t}$ posterior with great accuracy, including the regions toward its tails. We have chosen the KDE with Gaussian kernels for this proof-of-concept study because of its flexibility, but it may be worthwhile to experiment with distributions that can accommodate a positive skew, such as the skew normal distribution \citep{suyu2010dissecting, suyu2012cosmography}. 

\subsection{Computational efficiency}
\label{sec:computational_efficiency}
Computational efficiency is an important metric for the inference pipeline we present in this paper, because its use case lies in a joint-sample inference over many lenses. The total CPU time of our inference pipeline can be broken down into the data generation time, the BNN training time, and the $H_0$ inference time. Generating the training and validation sets consisting of 512,000 and 512 lenses, respectively, took a total of 6 hours on 8 CPU cores. Training the BNN with the architecture and training configuration, detailed in the Appendix (Section \ref{sec:appendix}), took less than 5 hours on a 16GB NVIDIA Tesla P100 GPU. 

The cosmological inference includes evaluating the BNN lens model on the test-set lenses and sampling from the $D_{\Delta t}^{(k)}$ posterior for individual lenses $k$. The former step can be done within seconds on the CPU or GPU across all the lenses at once. The latter step requires MCMC sampling, which dominates the total inference time. The median sampling time across 4 CPU cores was about 6 minutes per lens. The sampling time for a given lens depends on the shape of the caustic, which determines the stability of the lens equation solver called every iteration to solve for the image positions. Even for a fixed number of iterations, the MCMC sampling time varied greatly, from 3 to 50 minutes, across the lenses. Interestingly, of the eight lenses for which the MCMC sampling took more than 30 minutes, seven were doubles.

Taken together, the total computational time required to obtain the $H_0$ posterior from each test-set lens was 9 minutes. For 200 test-set lenses, this translates to around 30 hours -- which is promising, as complete experiments can be performed on shorter than a 1.25-day development cycle. Note also that the dataset generation and training time is a fixed investment that does not vary with the test set size. For 2000 test lenses, for instance, the pipeline would only take 6.3 minutes per lens.

It's useful to compare our computational efficiency with that of the traditional forward-modeling approach. Given the same set of simple model assumptions, \texttt{Lenstronomy} takes 0.1s per likelihood evaluation. It will yield a reasonable but unconverged $H_0$ posterior in 100,000 MCMC samples, or 3 hours, and will fully converge within 200,000 MCMC samples, or 6 hours. The 9-minute inference time of our method thus represents a speed-up of a factor of 20 to 40, with the range reflecting the degree of uncertainty in the \texttt{Lenstronomy} output.

\subsection{Limitations and Future Directions}
Having demonstrated proof of concept, we can now move onto testing the robustness of our approach to systematic errors, particularly those arising from a non-representative training set. We intend to apply the BNN hierarchical inference framework developed in \cite{wagner2020hierarchical} to the problem of inferring $H_0$ from large samples of lenses and recovering the population prior over the lens model parameters, in a follow-up study.

In this method paper, we did not test the response of the BNN to line-of-sight objects and image artifacts like cosmic rays. Such systematics tests must be performed before our method is applied to real data. One way to improve the BNN's robustness is augmenting the training data. \cite{hezaveh2017fast} added faint cosmic rays, hot pixels, and randomly distributed circular masks to their training images and reported good performance on real HST images.

The BNN lens modeling with lens light in the images is expected to improve with multi-band imaging, as demonstrated by experiments with non-Bayesian CNNs \citep{pearson2019use}. Encoding the color difference between the lens and source in the images will alleviate the contaminant effect of the deflector lens light. 

Another interesting avenue for exploration is advancing to more realistic sources. For the demonstrated method to be applicable to real systems, we must be able to handle more complex host galaxy profiles, including those of spiral galaxies, and possibly more than one source. The BNN can be put to the test of inferring the posterior over the coefficients of a shapelet decomposition \citep{birrer2016mass}.

Similarly, the lensing mass distribution can be made more complex. Whereas we have considered the total density profile in this work, we can probe the detailed structure of the lensing mass by adopting two-component models with explicitly assigned stellar and dark matter halo profiles. Disentangling the stellar and dark contributions would be particularly instructive for galaxy evolution studies. Other natural extensions include additional angular modes beyond elliptical symmetry and multi-plane, multi-deflector lensing.

As discussed in Section \ref{sec:additional_assumptions}, we have not addressed the internal and external mass sheets, which are potential sources of bias in $H_0$. These aspects will need to be investigated using separate datasets. The former can be probed with galaxy kinematics data, and the latter with photometry and spectroscopy of the environment. Our lens modeling pipeline thus requires an accompanying method that can characterize the mass sheets accurately from individual systems, so that biases can be mitigated hierarchically. The method must also be scalable, so as not to severely bottleneck the computation time.

While parameterizing the lens model posterior distribution as a mixture of two Gaussians served our simple profile assumptions, the training time and GPU memory requirement may not scale well to more complex lens and source profiles, given that the output dimension of the BNN increases exponentially with the number of model parameters.  More complex models may also demand more flexibility in the shape of the posterior. Likelihood-free inference methods such as flow-based generative models are likely to be interesting in that regard.

Lastly, our method can be applied to time delay cosmography with lensed supernovae (SNe) as well. The LSST is expected to discover 3,500 lensed SNe over the course of its 10-year survey \citep{goldstein2018rates}. With follow-up spectroscopy and time delay monitoring, the sample of ``cosmo-grade'' lensed SNe will be significantly increased and will benefit from the efficiency of BNN lens modeling.

\section{Conclusions}
\label{sec:conclusions}
In this paper, we introduced an automated pipeline for gravitational lens modeling and $H_0$ estimation that takes as input a high-resolution image, derives an approximate lens model parameter posterior PDF via a Bayesian neural network (BNN), and then propagates the posterior PDFs from multiple lenses into an estimate of the Hubble constant following the {H0LiCOW}/TDCOSMO project approach. The computational efficiency of our pipeline enables various sensitivity and robustness checks, from which we draw the following conclusions:
\begin{itemize}
    \item BNNs can yield accurate and well-calibrated posterior PDFs over the lens model parameters and the source position required for time delay cosmography.
    \item A simple combination of 200 mock test lenses yields a precision of 0.5 $\textrm{km s}^{-1} \textrm{ Mpc}^{-1}$ (0.7\%) and no detectable bias in $H_0$. For our choice of network architecture and and optimization strategy, the BNN lens modeling and the inferred $H_0$ are insensitive to varying image depth.
    \item Our inference pipeline takes around 9 minutes per lens, including the time taken to generate the training set, train the network, and run the cosmological sampling. It is automated and requires no expert supervision. This represents a 20-40x speed-up compared to the traditional forward modeling method. The computational efficiency makes the pipeline a promising method to handle large-scale sensitivity tests. 
\end{itemize}
The methodology and software presented in this paper promise to become core infrastructure in time delay cosmography, as the cosmology community prepares to beat down systematics for a large sample of lenses due to be available in a few years' time. 

The BNN-based $H_0$ inference pipeline presented in this paper provides a route to rapid inference of lens model parameters for a large sample of lenses. We have demonstrated that BNN lens modeling can accurately characterize the individual lens posterior PDFs and leads to an unbiased estimate of $H_0$ on a 200-lens sample, given simple assumptions on the lens model, time delay measurements, and the lens environment. The accuracy and speed make it a promising tool for the exploration of various systematics that may enter the $H_0$ analysis, where traditional forward modeling approaches could be slow and intractable. Given the large volumes of data expected from upcoming surveys, our pipeline can play a crucial role in time delay cosmography. 

\acknowledgments
\section*{Acknowledgments}
This paper has undergone internal review in the LSST Dark Energy Science Collaboration. We would like to thank the internal reviewers Thomas Collett and Remy Joseph for their insightful comments. 

JWP developed the \textsc{Baobab} and \textsc{H0rton} packages, used them to perform the analyses in this paper, and wrote the main text. 
SWC provided input on the BNN training and calibration, helped interpret the results, and contributed to the text. SWC was supported by the KIPAC-Chabolla fellowship and NSF Award DGE-1656518.
SB advised on motivation, scope, training set generation, and analysis and contributed to the text. 
JYL collaborated on the TDLMC submission by contributing code to the early versions of the image simulation and BNN training pipelines.
PJM advised on motivation, scope, experimental design, and analysis.
AR advised on scope and experimental design.

LSST DESC acknowledges ongoing support from the Institut National de Physique Nucl\'eaire et de Physique des Particules in France; the Science \& Technology Facilities Council in the United Kingdom; and the Department of Energy, the National Science Foundation, and the LSST Corporation in the United States. LSST DESC uses the resources of the IN2P3 / CNRS Computing Center (CC-IN2P3--Lyon/Villeurbanne - France) funded by the Centre National de la Recherche Scientifique; the Univ. Savoie Mont Blanc - CNRS/IN2P3 MUST computing center; the National Energy Research Scientific Computing Center, a DOE Office of Science User Facility supported by the Office of Science of the U.S.\ Department of Energy under contract No.\ DE-AC02-05CH11231; STFC DiRAC HPC Facilities, funded by UK BIS National E-infrastructure capital grants; and the UK particle physics grid, supported by the GridPP Collaboration.  This work was performed in part under DOE contract DE-AC02-76SF00515.

The authors thank the Google Cloud research credits program for providing the computing resources of the Google Cloud Platform. 

This work used the following public software packages: \textsc{H0rton} (this work), \textsc{Baobab} (this work), \textsc{Lenstronomy} \cite{birrer2018lenstronomy}, \textsc{emcee} \citep{foreman2013emcee}, \textsc{Corner} \citep{foreman2016corner}, \textsc{Astropy} \citep{robitaille2013astropy}, \textsc{Fastell} \citep{barkana1998fast}, and the standard Python libraries.  

\bibliography{main}

\section*{Appendix}
\label{sec:appendix}

\subsection{BNN implementation details} \label{sec:implementation_details}
\subsubsection{Scalability of the BNN with increasing target parameters} \label{sec:bnn_scalability}
More complex lens profiles will likely be described by more parameters. Yet the size of the output dimension scales exponentially with the number of model parameters, an oft-criticized trait of BNNs. This effect can be mitigated by parameterizing the covariance matrix as positive diagonal elements plus a low-rank (rank $r$) matrix, in which case the $p_{\rm out}=2\times (2 + r)p + 1$ and the scaling is instead linear. See the \texttt{LowRankGaussianNLL} and \texttt{DoubleLowRankGaussianNLL} classes in the \textsc{H0rton} repo for the implementation.

\subsubsection{Numerical stability of the loss function}
As presented in Section \ref{sec:posterior_inference}, the ELBO objective is the negative log of the posterior given in Equation \ref{eq:aleatoric_posterior} plus an $L_2$ weight regularization term with strength $\lambda$. Given our double Gaussian parameter posterior assumption, this can be written more concretely as:
\begin{align} \label{eq:bnn_loss_implemented}
&\mathcal{L}(W) = -\log\left(w_1(d; W)\right) - \log(1 - w_1(d; W)) \nonumber \\
& + \frac{1}{2}\log |\Sigma_1(d; W)| + \frac{1}{2}\log |\Sigma_2(d; W)| \nonumber \\
& + \left(m_1(d; W) - \mu_1\right)^T \Sigma_1(d; W)^{-1} \left(m_1(d; W) - \mu_1\right)  \nonumber \\ 
& + \left(m_2(d; W) - \mu_2\right)^T \Sigma_2(d; W)^{-1} \left(m_2(d; W) - \mu_2\right) \nonumber \\
& + \lambda ||W||^2
\end{align}
where $m_1, m_2$ represent the network-predicted posterior means for the two Gaussians. The dependence of each posterior parameter on $W$ and the input training image $d$ is made explicit here. To ensure that the optimization is numerically stable and well-defined, each covariance matrix was parameterized as the log Cholesky decomposition of its inverse (the precision matrix), i.e. for $\Sigma(d; W) = \Sigma_{1/2}(d; W)$,
\begin{align}
&\Sigma^{-1}(d; W) = L(d; W) L(d; W)^T \nonumber \\
 &  L(d; W) = \begin{bmatrix} \exp l_{1}(d; W) & 0  & 0 \\
L_{21}(d; W) & \exp l_{2}(d; W) & 0\\ L_{31}(d; W) & L_{32}(d; W) & \exp l_{3}(d; W)\end{bmatrix}. 
\end{align}
Note that the BNN predicts the log of the diagonal entries, so that the diagonal entries are positive. This extra requirement of the log Cholesky parameterization guarantees that $\Sigma$ is positive definite, whereas the regular Cholesky parameterization only guarantees that $\Sigma$ is positive semidefinite and can thus lead to a non-unique $L$. Also, we parameterized $w_1$ as the half-sigmoid of the BNN-predicted logit $\omega$ to get it in the range $\left(0, \frac{1}{2} \right)$.
\begin{align} \label{eq:sigmoid}
w_1(d;W) = \sigma \left(\omega(d; W) \right) \equiv \frac{1}{1 - \exp(-\omega(d;W)) }
\end{align}
The source position $x_{\rm src}, y_{\rm src}$ were parameterized in terms of their offsets from the lens position $x_{\rm lens}, y_{\rm lens}$.

\subsubsection{Deep residual networks with Monte Carlo dropout} \label{sec:resnet}
ResNets, or deep residual networks, address the problems of vanishing/exploding gradients \citep{bengio1994learning} and degradation of training accuracy known to plague deep networks. They do so by inserting so-called ``shortcut connections'' between the inputs and outputs of a few stacked convolutional layers \citep{he2015deep}. The idea is that, instead of expecting a set of stacked layers to learn the mapping $H$ between the input $x$ and output $H(x)$, we require it to learn the difference, i.e. the residual mapping $F(x) \equiv H(x) - x$. The original mapping is then recovered as $F(x) + x$. The shortcut connections implement precisely this addition operation. See Figure \ref{fig:network_architecture} for the architecture of \texttt{ResNet101} as applied to our images. We inserted 1D dropout before every convolution, including the first convolution prior to max-pooling. We also had batch normalization and ReLU activation after every convolution. \texttt{ResNet101} has 44 million trainable parameters and 347 layers.

The depth and width of the architecture were also tunable hyperparameters. We chose \texttt{ResNet101} among ResNet variants with different depths and widths -- the other candidates being \texttt{ResNet50}, which was shallower but equally wide, and \texttt{ResNet56}, which was shallower but much wider -- based on the validation-set performance.

\end{document}